\documentclass[showpacs,preprintnumbers,twocolumn,nofootinbib,eqsecnum,prd]{revtex4-1}  

\usepackage{graphicx}
\usepackage{latexsym}
\usepackage{amsmath}
\usepackage{amssymb}

\def\beq{\begin{equation}}
\def\eeq{\end{equation}}
\def\pmb#1{\setbox0=\hbox{$#1$}%
  \kern-.025em\copy0\kern-\wd0
  \kern.05em\copy0\kern-\wd0
  \kern-.025em\raise.0433em\box0}
\def\bfOmega{\pmb{\Omega}}

\begin{document}

\title{Two-body gravitational spin-orbit interaction at linear order in the mass ratio}

\author{Donato \surname{Bini}$^1$}
\author{Thibault \surname{Damour}$^2$}

\affiliation{$^1$Istituto per le Applicazioni del Calcolo ``M. Picone'', CNR, I-00185 Rome, Italy\\
$^2$Institut des Hautes Etudes Scientifiques, 91440 Bures-sur-Yvette, France}

\date{\today}

\begin{abstract}
We analytically compute, to linear order in the mass-ratio, the \lq\lq geodetic" spin precession frequency of a small spinning body orbiting a large (non-spinning) body to
the eight-and-a-half post-Newtonian order, thereby extending previous analytical knowledge which was limited to the third post-Newtonian level. These results are obtained applying analytical gravitational self-force theory  to the first-derivative level generalization of Detweiler's gauge-invariant redshift variable.
We compare our analytic results with  strong-field numerical data recently obtained by S.~R.~Dolan et al. [Phys.\ Rev.\ D {\bf 89}, 064011 (2014)].
Our new, high-post-Newtonian-order results capture the strong-field features exhibited by the  numerical data.
We argue that the spin-precession will diverge as $\approx -0.14/(1-3y)$ as the light-ring is approached.
We transcribe our kinematical spin-precession results into a corresponding  improved analytic knowledge of one of the two (gauge-invariant) effective gyro-gravitomagnetic ratios characterizing spin-orbit couplings within the effective-one-body formalism.
We provide simple, accurate analytic fits both for spin-precession and  the effective gyro-gravitomagnetic ratio. The latter fit predicts that the linear-in-mass-ratio correction to the gyro-gravitomagnetic ratio changes sign before reaching the light-ring.
This strong-field prediction might be important  for improving the analytic  modeling of coalescing spinning binaries.
\end{abstract}
\maketitle

\section{Introduction}

The general relativistic two-body problem has become of renewed importance in view of the current development of gravitational-wave detectors. It is plausible that the first detections will concern binary systems made of {\it spinning} black holes, because the spin-orbit interaction can increase the binding energy of the last stable (circular) orbit, and thereby lead to a larger gravitational wave output.
This gives a strong incentive for improving our analytic knowledge of the gravitational spin-orbit interaction in binary systems.
Several analytical-relativity methods are currently being actively pursued: post-Newtonian (PN) theory \cite{Schafer:2009dq,Blanchet:2013haa}, gravitational self-force (GSF) theory in black hole backgrounds \cite{Poisson:2011nh,Barack:2009ux}, and the effective-one-body (EOB) formalism \cite{Buonanno:1998gg, Buonanno:2000ef,Damour:2000we,Damour:2001tu}.
Recent years have witnessed a useful synergy between these methods, with information flowing between them (and flowing also from the results of full numerical relativity simulations) \cite{Detweiler:2008ft,Damour:2009sm,Blanchet:2009sd,Blanchet:2010zd,Barack:2010ny,LeTiec:2011ab,LeTiec:2011dp,Barausse:2011dq,Shah:2013uya,LeTiec:2011bk,Damour:2011fu,Akcay:2012ea,Hinder:2013oqa,Bini:2014nfa,Damour:2014afa,Dolan:2013roa}.

The present work will show another instance of such a synergy between analytical methods: we consider a binary system where a body having a small mass $m_1$ and a small spin ${\mathbf S}_1$ moves on a circular orbit of radius $r_0$ and orbital frequency $\Omega$ around a non-spinning body having a large mass $m_2$.
We shall start by using analytic GSF theory \cite{Mano:1996vt,Mano:1996mf,Mano:1996gn,Barack:1999wf, Barack:2001gx,Hikida:2004hs,Hikida:2004jw} to compute to linear order in the mass ratio $m_1/m_2$, the spin-orbit coupling ${\bfOmega}_1^{\rm SO}\cdot {\mathbf S}_1$ 
of particle 1, which is equivalent to knowing the spin precession ${\bfOmega}_1^{\rm SO}$ of ${\mathbf S}_1$ \cite{Damour:2007nc}.

The GSF result for ${\bfOmega}_1^{\rm SO}$ (considered as a function of $\Omega$) will be expanded to high order in a PN expansion, by using the same technology that we recently used \cite{Bini:2014nfa,Bini:2013zaa,Bini:2013rfa}
to derive the PN expansion of Detweiler's redshift function \cite{Detweiler:2008ft}.
The latter PN-expanded result will then be transcribed within the EOB formalism in terms of higher order contributions to  one of the two (gauge-invariant) effective gyro-gravitomagnetic ratios ($g_{S^*}^{\rm eff}$) characterizing spin-orbit couplings within the EOB formalism.
In addition, we shall compare our results to a recent work by Dolan et al. \cite{Dolan:2013roa} which has provided both an analytical GSF derivation of ${\bfOmega}_1^{\rm SO}$ and numerical data on the strong-field behavior of the first-order GSF contribution to ${\bfOmega}_1^{\rm SO}$.
We shall find that our new high PN-order results on ${\bfOmega}_1^{\rm SO}$ are able to capture strong-field features that are beyond the reach of the currently available low PN-order ones.

Detweiler \cite{Detweiler:2008ft} has emphasized the usefulness of computing gauge-invariant quantities from GSF theory.
He pointed out that (quasi-)circular orbits provide a calculable gauge-invariant function $|k|(\Omega)$ via the existence of a helical Killing vector
\beq
\label{1.1}
k^\mu \partial_\mu=\partial_t +\Omega \partial_\phi\,.
\eeq
The function $|k|(\Omega)$ relates the norm $|k|$ of $k^\mu$ [computed with the regularized perturbed metric $g_{\mu\nu}^{\rm R}(x^\lambda)$ \cite{Detweiler:2000gt,Detweiler:2002mi},  
and evaluated along the world line 
${\mathcal L}_1$ of particle 1],
\beq
\label{1.2}
|k|^2=-[g_{\mu\nu}^{\rm R} k^\mu k^\nu]_1\,,
\eeq
to the orbital frequency $\Omega$ of the considered circular motion.
The bracket notation $[\ldots ]_1$ in Eq. (\ref{1.2}) refers to an evaluation along the world line 
${\mathcal L}_1$ of particle 1.

Along the same vein, the existence of a Killing vector field $k^\mu (x^\lambda)$ implies the existence of further gauge-invariant functions. Indeed, at the one-derivative level, the (helical) world line of particle 1 is endowed with the following geometrical configuration: $\{[k^\mu]_1, [K_{\mu\nu}]_1\}$. This geometrical configuration is defined by evaluating on 
${\mathcal L}_1$ the vector field $k^\mu (x^\lambda)$ together with the tensor field
\beq
\label{1.3}
K_{\mu\nu}(x^\lambda)\equiv \nabla_\mu^{\rm R}k_\nu(x^\lambda)\,,
\eeq
where $\nabla_\mu^{\rm R}$ denotes the covariant derivative associated with the regularized metric $g_{\mu\nu}^{\rm R}(x^\lambda)$.
As $k^\mu$ is a Killing vector of $g_{\mu\nu}^{\rm R}$, we have
\beq
\label{1.4}
K_{\mu\nu}=\nabla_\mu^{\rm R} k_\nu=-\nabla_\nu^{\rm R} k_\mu=\frac12 (\partial_\mu k_\nu-\partial_\nu k_\mu)\,.
\eeq
Moreover, as $[k^\mu]_1$ differs from the unit four-vector $u^\mu$ tangent to ${\mathcal L}_1$ only by the constant factor $|k|$, and as ${\mathcal L}_1$ is a geodesic
of $g_{\mu\nu}^{\rm R}$ \cite{Detweiler:2000gt,Detweiler:2002mi}, we have
\beq
\label{1.5}
[k^\lambda \nabla_\lambda^{\rm R}k_\mu]_1=[k^\lambda K_{\lambda\mu}]_1=0\,.
\eeq
If we view the antisymmetric tensor $K_{\mu\nu}$ as an electromagnetic field, Eq. (\ref{1.5}) says that its electric-vector projection
$E_\mu\equiv [K_{\mu\nu}u^\nu]_1=[K_{\mu\nu}k^\nu/|k|]_1$ vanishes. As a consequence, the geometrical configuration $\{[k^\mu]_1, [K_{\mu\nu}]_1\}$ is equivalent to the configuration defined by the two four-vectors $\{[k^\mu]_1, B_\mu \}$, where
\beq
\label{1.6}
B_\mu =  [K^*_{\mu\nu}u^\nu]_1=  \left[ K^*_{\mu\nu}\frac{k^\nu}{|k|} \right]_1\,,
\eeq
with $K^*_{\mu\nu}\equiv \frac12 \eta_{\mu\nu\alpha\beta}K^{\alpha\beta}$ denoting the Hodge-dual of the two-form $K_{\mu\nu}$.
[Here, and below, all the Riemannian constructs are defined by the regularized perturbed metric $g_{\mu\nu}^{\rm R}$; e.g., $K^{\alpha\beta}\equiv g_{\rm R}^{\alpha\mu}g_{\rm R}^{\beta\nu}K_{\mu\nu}$; $\eta_{\mu\nu\alpha\beta}=\sqrt{-g_{\rm R}}\, \epsilon_{\mu\nu\alpha\beta}$ with $\epsilon_{0123}=+1$.]

As the two vectors $[k^\mu]_1$ and $B_\mu$ are orthogonal , the only invariants defined by the configuration $\{[k^\mu]_1, B_\mu \}$ are the two scalars measuring their norms, ie., $|k|$, Eq. (\ref{1.2}) and $|\nabla k|$, defined by
\beq
|\nabla k|^2\equiv [g_{\mu\nu}^{\rm R} B^\mu B^\nu]_1=\frac12 [K_{\mu\nu}K^{\mu\nu}]_1\,.
\eeq 
Therefore, at the one-derivative level, there is only one new gauge-invariant function associated with $\{k^\mu, \nabla_\mu k_\nu\}$,  namely the function $|\nabla k|(\Omega)$ relating the norm $|\nabla k|$ of $\nabla_\mu k_\nu$ to $\Omega$.

At the two-derivative level, we have, along the world line 
${\mathcal L}_1$, a geometric configuration defined by $[k^\mu]_1$, $[\nabla_\mu^{\rm R}k_\nu]_1$ and $[\nabla_{\mu\nu}^{\rm R}k_\lambda]_1$ together with the curvature tensor $R^{\rm R}_{\mu\nu\rho\sigma}$ of $g_{\mu\nu}^{\rm R}$. In view of the consequence $\nabla^{\rm R}_{\mu\nu}k_\lambda=R^{\rm R}_{\lambda\nu\mu\sigma}k^\sigma $
of Killing's equation, and of the Ricci-flatness of  $g_{\mu\nu}^{\rm R}$ \cite{Detweiler:2002mi}, such a configuration is equivalent to the configuration $\{[k^\mu]_1, B_\mu, {\mathcal E}_{\mu\nu}, {\mathcal B}_{\mu\nu}\}$, where $({\mathcal E}_{\mu\nu}, {\mathcal B}_{\mu\nu})$ denote the electric-like and magnetic-like projections of the Riemann (or, equivalently, Weyl) tensor of $g_{\mu\nu}^{\rm R}$ with respect to the time direction $u^\mu=k^\mu/|k|$; e.g., \beq
{\mathcal E}_{\mu\nu}\equiv [R^{\rm R}_{\mu\alpha\nu\beta}u^\alpha u^\beta]_1\,.
\eeq

By contrast to the one-derivative level where one could only extract one geometric scalar  (beyond the zero-derivative scalar $|k|$), it is clear that the two-derivative level gives access, in general, to several new scalars (notably the four independent eigenvalues of the traceless symmetric tensors ${\mathcal E}_{\mu\nu}$ and  ${\mathcal B}_{\mu\nu}$, as well as the other scalars one can build by combining $B^\mu$ with ${\mathcal E}_{\mu\nu}$ and  ${\mathcal B}_{\mu\nu}$). 
One of the simplest of those scalars is ${\mathcal E}^2\equiv {\mathcal E}_{\mu\nu}{\mathcal E}^{\mu\nu}$.
Correspondingly, one can extract several gauge-invariant functions of $\Omega$, such as ${\mathcal E}^2(\Omega)$. See below for more details.

The Detweiler-like gauge-invariant functions $|k|(\Omega)$, $|\nabla k|(\Omega)$, ${\mathcal E}^2(\Omega), \ldots$, discussed above have, a priori, only a {\it kinematical}
significance. However, they turn out to have also a {\it dynamical} significance. In the case of the original redshift function $u^t_1(\Omega)=dt/ds_1=1/|k|(\Omega)$, its dynamical meaning was discovered in Refs. \cite{LeTiec:2011ab,LeTiec:2011dp,Barausse:2011dq}. 

In particular the first-order self-force (1SF) contribution to $|k|^2(\Omega)$, i.e., essentially, the double contraction
\beq
\label{1.8}
h_{kk}(\Omega)\equiv [k^\mu k^\nu h^{\rm R}_{\mu\nu}(x^\lambda)]_1\,,
\eeq
where $h^{\rm R}_{\mu\nu}$ denotes the regularized version of the mass-ratio-rescaled 1SF perturbation $h_{\mu\nu}$ of the background metric (here taken to be a Schwarzschild metric of mass $m_2$),
\beq
\label{1.9}
g_{\mu\nu}(x^\lambda; m_1, m_2)=g^{(0)}_{\mu\nu}(x^\lambda;  m_2)+\frac{m_1}{m_2}h_{\mu\nu}(x^\lambda)+O\left(\frac{m_1^2}{m_2^2}\right)\,,
\eeq
was found to be very simply related to the 1SF contribution $a^{1\rm SF}(u)$ to the main EOB potential $A(u;\nu)$ \cite{Buonanno:1998gg, Buonanno:2000ef,Damour:2000we} describing the interaction energy of the two masses $m_1$, $m_2$
\beq
\label{1.10}
A(u;\nu)=1-2u+\nu a^{1\rm SF}(u) +O(\nu^2)\,,
\eeq
(where $\nu\equiv m_1m_2/(m_1+m_2)^2$ is the symmetric mass ratio) namely \cite{Barausse:2011dq,Akcay:2012ea}
\beq
\label{1.11}
a^{1\rm SF}(u)=-\frac12 h_{kk}-\frac{u(1-4u)}{\sqrt{1-3u}}\,.
\eeq

At the one-derivative level, the quantity $|\nabla k|$ has the kinematical meaning of the relative precession frequency between parallely propagated and Lie-dragged spatial frames along the world line ${\mathcal L}_1$ \cite{Iyer:1993qa,Dolan:2013roa}. As noted in \cite{Iyer:1993qa}, if the Killing vector $k^\mu$ were simply
$k^\mu_0\partial_\mu=\partial_t$, $|\nabla k_0|$ would measure the gyroscopic precession of the (parallely propagated)  spin vector $S_1^\mu$ with respect to fixed stars at infinity.
In our case, $k_0^\mu \partial_\mu=\partial_{t}$, is not a Killing vector, i.e., it does not generate a continuous one-parameter isometry group; however, it generates a {\it discrete isometry} group (made of $g_n={\rm exp}(nP\partial_{t})$,  where $n$ is an  integer and $P\equiv 2\pi/\Omega$ the orbital period). This fact implies that, after each orbital period, the difference $||\nabla k|-\Omega |$ (where $\Omega$ subtracts the effect of the $\Omega$-rotation contained in the $\Omega \partial_\phi$ piece in $k^\mu$) measures, in a stroboscopic way, the precession of $S_1^\mu$ with respect to fixed stars at infinity (i.e., with respect to $\partial_t$ at infinity). More precisely, as one finds that the Lie-dragged frame lags behind the parallely propagated frame (i.e., that $|\nabla k|$ is smaller than $\Omega$), the stroboscopic spin precession is measured (in a frame comoving with particle 1) by the angular frequency
\beq
\label{1.12}
\Omega_{\rm prec}=\Omega-|\nabla k|\,.
\eeq   
The dynamical meaning of this gauge-invariant kinematical quantity is furnished by the results of Ref. \cite{Damour:2000we}, which related the spin angular frequency (as seen in a $3+1$ Hamiltonian description) to the spin-orbit coupling terms $\bfOmega_1^{\rm SO} \cdot {\mathbf S}_1+\bfOmega_2^{\rm SO} \cdot {\mathbf S}_2$
(with $\bfOmega_1^{\rm SO}$ and $\bfOmega_2^{\rm SO}$ proportional to ${\mathbf L}={\mathbf x}\times {\mathbf p}$) in the Hamiltonian. As we shall discuss in detail below, in the special case of circular orbits we have the simple numerical link
\beq
\label{1.13}
\Omega_1^{\rm SO}=\Omega_{\rm prec}=\Omega-|\nabla k|\,,
\eeq
in spite of the fact that $\Omega_1^{\rm SO}$ and $\Omega_{\rm prec}$ a priori measure spin precessions in different frames ($\Omega_{\rm prec}$ referring to a comoving frame linked to $u^\mu$, while $\Omega_1^{\rm SO}$ refers to a laboratory frame linked to $\partial_t$). The simple link (\ref{1.13}) along circular orbits has also been recently pointed out in Ref. \cite{Dolan:2013roa}.

At the two-derivative level, some of the gauge-invariant kinematical functions that one can compute have also an interesting dynamical significance. For instance, the quantity $\int \mu_1 {\mathcal E}^2 ds_1=\int \mu_1 {\mathcal E}_{\mu\nu}{\mathcal E}^{\mu\nu} ds_1$ describes the leading-order tidal correction to the effective action of the binary system associated with the quadrupolar tidal deformation of body 1 \cite{Damour:2009wj}. [Here, $\mu_1$ is the quadrupolar tidal polarization parameter of body 1]. Recently, the regularized invariant ${\mathcal E}^2$ has been computed in PN theory at the second post-Newtonian level \cite{Bini:2012gu}. We see that, in principle, GSF theory could compute it to all PN orders, but to first order in $m_1/m_2$.

Leaving to future work such a GSF study of higher-derivative level gauge-invariant quantities, we focus in the present paper on the spin-orbit function $\Omega_1^{\rm SO}(\Omega)$ and on its EOB reformulation in terms of the gauge-invariant gyro-gravitomagnetic ratio $g_{S^*}^{\rm eff}$.

\section{Spin-orbit coupling and its link with spin precession}

In this section we start by considering a general binary system, with arbitrary masses $m_1$ and $m_2$ and spins ${\mathbf S}_1$ and ${\mathbf S}_2$.
Before studying the GSF limit where $m_1\ll m_2$, let us recall that, within PN theory, the motion of each member of such a binary system is described, to high PN accuracy, by the usual (Mathisson-Papapetrou) equations of motion valid for a {\it test} particle in an external metric $g_{\mu\nu}^{\rm ext}$, except for the fact that
the values of $g_{\mu\nu}^{\rm ext}$ and its derivatives appearing in these equations have to be replaced by the {\it regularized} values of the full two-body metric
$g_{\mu\nu}(x^\lambda; m_1, m_2, {\mathbf S}_1,{\mathbf S}_2)$ (see, e.g., \cite{Damour:1982wm}). 
The most efficient, and consistent, regularization method for this purpose has been found to be dimensional regularization. An analog of this fact has been proven to hold within GSF theory \cite{Detweiler:2000gt,Detweiler:2002mi}, with the bonus that, at the first GSF order, the regularized values of $g_{\mu\nu}(x^\lambda)$, $\partial_\sigma g_{\mu\nu}(x^\lambda), \ldots$ can be obtained by evaluating on the world line ${\mathcal L}_1$
of body 1, the values of a regular metric $g_{\mu\nu}^{\rm R}(x^\lambda)$ and of its derivatives   $\partial_\sigma g_{\mu\nu}^{\rm R}(x^\lambda), \ldots$.
In addition, the regular metric  $g_{\mu\nu}^{\rm R}(x^\lambda)$ satisfies (to linear order) Einstein's vacuum equations. [The PN analog of this property being that the regularized value on ${\mathcal L}_1$ of $R_{\mu\nu}(g_{\mu\nu}^{\rm PN}(x))$ vanishes.]

If we work to linear order in ${\mathbf S}_1$, we have, either in PN or GSF theory, equations of motion of the type
(with a superscript ${\rm R}$ denoting regularization)
\begin{eqnarray}
\label{2.1}
\frac{D^{\rm R}}{ds_1}(m_1u^\mu)&=& -\frac12 R^{\rm R}{}^\mu{}_{\nu\alpha\beta}u^\nu S_1^{\alpha\beta}+O(S_1^2)\,,\\
\label{2.2}
\frac{D^{\rm R}}{ds_1}(S_1^\mu)&=&O(S_1^2)\,,
\end{eqnarray}
where ${D^{\rm R}}/{ds_1}=u^\mu \nabla_\mu^{\rm R}$, with $u^\mu=dx_1^\mu/ds_1$  the unit tangent to ${\mathcal L}_1$, where $S_1^{\alpha\beta}\equiv\eta^{\alpha\beta\mu\nu}u_\mu S_1{}_\nu$, and where we used the spin supplementary condition $S_1^{\alpha\beta}p_\beta=0$ (which yields $S_1^{\alpha\beta}u_\beta=O(S_1^2)$ and $p_\alpha=m_1 u_\alpha +O(S_1^2)$).

Note that, when working to linear order in ${\mathbf S}_1$, we must a priori keep the curvature-spin force appearing on the right hand side of Eq. (\ref{2.1}).
In other words, ${\mathcal L}_1$ is not a geodesic of $g_{\mu\nu}^{\rm R}(x^\lambda)$. However, as emphasized in \cite{Damour:2000we}, if our aim is to derive the spin-orbit coupling term $ {\bfOmega}_1^{\rm SO}\cdot {\mathbf S}_1$ in the Hamiltonian, the crucial equation is Eq. (\ref{2.2}), in which we can consistently neglect the $O(S_1)$ contributions both to $g_{\mu\nu}^{\rm R}(x^\lambda)$ and to the evolution of the world line ${\mathcal L}_1$. In that sense, we can analyze the consequences of Eq. (\ref{2.2}) assuming that ${\mathcal L}_1$ is a geodesic of $g_{\mu\nu}^{\rm R}(x^\lambda)$ [i.e., neglecting the force on the right hand side of Eq. (\ref{2.1})].

Besides the simplifications brought by neglecting the contribution of order $O(S_1^2)$, we can further simplify the derivation of the spin-orbit coupling,   and of its link with $|\nabla k|$, by noting that (when using polar-type coordinates
 $t,r,\theta ,\phi$)  the motion will stay within the equatorial plane $\theta=\pi/2$ if it starts within it (here we assume that the vector ${\mathbf S}_2$ is orthogonal to the orbital plane). In mathematical terms, this means that the (2+1)-dimensional hypersurface $\theta=\pi/2$ is totally geodesic, so that, for analyzing the consequences of the geodesic equation $D^{\rm R} u^\mu/ds_1=0$, and of the parallel transport equation (\ref{2.2}), it is enough to work with the (2+1)-dimensional metric restricted to $\theta=\pi/2$. Using the coordinates
\beq
\bar t\equiv t \,,\quad \bar \phi\equiv  \phi-\Omega t\,,
\eeq
adapted to the helical symmetry
\beq
\label{2.3}
k^\mu \partial_\mu =\partial_t +\Omega \partial_\phi=\partial_{\bar t}\,,
\eeq
the general structure of the equatorially-reduced (2+1)-dimensional metric reads (suppressing, for brevity, the superscript R)
\beq
\label{2.4}
ds^2_{(2+1)}=g^{(2+1)}_{\bar \mu  \bar \nu}dx^{\bar \mu}dx^{\bar \nu}\,,\quad (\bar \mu, \bar \nu=0,1,2)\,,
\eeq
where, say, $x^0=\bar t$, $x^1=r$, $x^2=\bar \phi$. In this coordinate system, the metric is stationary, 
\beq
\label{2.4bis}
\partial_{\bar t}g_{\bar \mu  \bar \nu}=0,
\eeq 
and the world line ${\mathcal L}_1$ is \lq\lq vertical": $u^\mu \propto \delta_{\bar t}^\mu$.
The geodesic condition for ${\mathcal L}_1$ then implies that the Christoffel symbol $\Gamma^{\bar i}_{\bar t \bar t}=0$, that is 
\beq
\label{2.5}
\partial_{\bar i}g_{\bar t \bar t}=0\,.
\eeq
This is equivalent to the condition $0=k^\lambda \nabla_\lambda k_\mu=-
k^\lambda \nabla_\mu k_\lambda=-\frac12 \nabla_\mu k^2$ mentioned in the Introduction.
The spin condition $u^\mu S_\mu=0$ (where we provisionally omit the body label 1 on $S_1^\mu$) then says that $S_{\bar t}=0$. The two remaining covariant components of the spin vector
$S_{\bar i}=(S_1,S_2)=(S_r,S_{\bar \phi})$ then satisfy
\beq
\label{2.6}
\frac{dS_{\bar i}}{dt}=\Gamma^{\bar j}{}_{\bar i \bar t}S_{\bar j}\,,
\eeq
which can be rewritten as
\beq
\label{2.6bis}
\frac{dS_{\bar i}}{dt}=K_{\bar i}{}^{\bar j} S_{\bar j}\,, 
\eeq
because we have (using $\partial_{\bar t} g_{\bar i\bar j}=0$)
\beq
\label{2.7}
\Gamma^{\bar j}{}_{\bar i \bar t}=\frac12 g^{\bar j \bar k}(\partial_{\bar i}g_{\bar t\bar k}-\partial_{\bar k}g_{\bar t\bar i})
\equiv  g^{\bar j \bar k} K_{\bar i \bar k}\,.
\eeq
Here, we used the fact that the covariant components $k_{\bar \mu}$ of $k=\partial_{\bar t}$ are simply $k_{\bar \mu}=g_{\bar \mu \bar t}$, so that the covariant components of $K_{\bar \mu \bar \nu}$, Eq. (\ref{1.4}), are $K_{\bar i \bar t}=\frac12 \partial_{\bar i}g_{\bar t \bar t}=0$ [in view of Eq. (\ref{2.5})], and $K_{\bar i \bar j}=\frac12 (\partial_{\bar i}g_{\bar j \bar t}-\partial_{\bar j}g_{\bar i \bar t})$.

The evolution equation (\ref{2.6bis}) for the nonvanishing covariant components of the spin vector $S_{\bar \mu}=(0,S_{\bar i})$, together with the antisymmetry of $K_{\bar i \bar j}$ and the stationarity of the metric (\ref{2.4bis}), implies that $g^{\bar i \bar j}S_{\bar i}S_{\bar j}=s^2$ remains constant as $S_{\bar i}$ evolves.
If we introduce a $\bar t$-independent\footnote{In geometrical terms, the $\bar t$-independence of the spatial frame, $\partial_{\bar t} e_{\hat a}^{\bar i}=0 $, means that its Lie-derivative along $k=\partial_{\bar t}$ vanishes, i.e., that it is Lie-dragged along $k$.} spatial frame $e_{\hat a}^{\bar i}\partial_{\bar i}$ such that $g^{\bar i\bar j}=\delta^{\hat a \hat b}e_{\hat a}^{\bar i}e_{\hat b}^{\bar j}$ (a convenient particular choice for defining $e_{\hat a}^{\bar i}$ will be discussed below), the frame components
$S_{\hat a}=S_{\bar i}e_{\hat a}^{\bar i}$ of the spin vector will evolve as
\beq
\label{2.9}
\frac{dS_{\hat a}}{dt}=K_{\hat a \hat b}S_{\hat b}\,.
\eeq
In terms of those frame components the conservation of $g^{\bar i \bar j}S_{\bar i}S_{\bar j}=s^2$ leads to the conservation of 
the Euclidean norm $S_{\hat a}S_{\hat a}=s^2$. Actually, as the indices ${\hat a}$, ${\hat b}$ take only two values, $K_{\hat a \hat b}$ has only one independent component $K_{\hat 1 \hat 2}=-K_{\hat 2 \hat 1}$, and Eq. (\ref{2.9}) describes a rotation of the Euclidean two-vector $(S_{\hat 1},S_{\hat 2})$ with angular frequency $-K_{\hat 1 \hat 2}$. In other words, the precession of $S_{\hat a}$ is associated with the unique
independent component of the tensor $K_{\hat a \hat b}$ and can therefore be conveniently computed (in any frame or coordinate system) from
\beq
\label{2.10}
|\nabla k|^2=\frac12 K_{\mu\nu}K^{\mu\nu}=K_{\hat 1 \hat 2}^2\,.
\eeq
Eqs. (\ref{2.9}) and (\ref{2.10}) are equivalent to results of \cite{Dolan:2013roa}.
This precession is also conveniently encoded in the dual of the three-form $k^\flat \wedge dk^\flat$ (where $k^\flat=k_\mu dx^\mu$ is the one-form associated with $k^\mu$), namely
\begin{eqnarray}
\label{2.11}
|\nabla k |&=&   
\frac12 \eta^{\mu\nu\lambda}\frac{k_\mu}{|k|} (\partial_\nu k_\lambda -\partial_\lambda k_\nu)\nonumber\\
&=&  
\frac{\sqrt{-g_{\bar t \bar t}}}{\sqrt{-g^{(2+1)}}}(\partial_r g_{\bar \phi \bar t}-\partial_{\bar \phi} g_{r \bar t})\,,
\end{eqnarray}
where we used $k_{\bar t}=g_{\bar t \bar  t}$, $|k|=\sqrt{-g_{\bar t \bar t}}$, and, for the $(2+1)$-dimensional Levi-Civita contravariant tensor, $\eta^{012}=-1/\sqrt{-g^{(2+1)}}$. [We shall check below that the right hand side of Eq. (\ref{2.11}) is positive].
The latter formula  is technically useful because it allows one to  express
$|\nabla k|$ entirely in terms of the covariant components of the (regularized) $(2+1)$-dimensional metric $g_{\mu\nu}$ in the original non-comoving coordinates $t,r,\phi$. Indeed, it suffices to replace
$t=\bar t$, $\phi=\bar \phi +\Omega \bar t$ in $ds^2=g_{\mu\nu}dx^\mu dx^\nu$ to get the components $g_{\bar \mu \bar \nu}dx^{\bar \mu} dx^{\bar \nu}$ associated with the comoving coordinates $\bar t$, $r$ and $\bar \phi$. For instance,
\begin{eqnarray}
\label{2.12}
g_{\bar t \bar t}&=& g_{\mu\nu}k^\mu k^\nu=g_{00}+2\Omega g_{0\phi}+\Omega^2 g_{\phi\phi}\nonumber\\
\label{2.13}
g_{\bar t \bar i}&=& g_{ i \nu}k^\nu=g_{ i 0}+\Omega g_{  i \phi}\qquad (i=r, \phi)\,.
\end{eqnarray}
Moreover, as the Jacobian of the transformation $(t,r,\phi) \to (\bar t,r,\bar \phi) $ is equal to one, the determinant appearing in the denominator of Eq. (\ref{2.11}) 
can be simply computed as ${\rm det}\, (g_{\mu\nu}^{(2+1)})$, with the original components $g_{\mu\nu}^{(2+1)}$ of the equatorial metric in non-comoving $t,r,\phi$ coordinates.

The result (\ref{2.9}), with $K_{\hat 1 \hat 2}=|\nabla k|$, concerns the geometrical precession, with respect to a spatial frame orthogonal to ${\mathcal L}_1$ and Lie-dragged along it, of a local spin vector orthogonal to ${\mathcal L}_1$. One can relate this kinematical fact to dynamical properties of spin-orbit coupling in a binary system by using results of Ref. \cite{Damour:2007nc}. 
Some delicate aspects of the connection between the local spin-precession (\ref{2.9}) and the spin-orbit piece, $H_{\rm SO}=\bfOmega_1^{\rm SO}\cdot {\mathbf S}_1^{\rm can}+\bfOmega_2^{\rm SO}\cdot {\mathbf S}_2^{\rm can}$, in the Hamiltonian of a binary system have to be noted (besides \cite{Damour:2007nc}, see also Refs. \cite{Khriplovich:2008ni,Barausse:2009aa}). First, Ref. \cite{Damour:2007nc} pointed out that the usual Poisson brackets of the Cartesian components of a (constant magnitude, i.e., $(S_{1x}^{\rm can})^2+(S_{1y}^{\rm can})^2+(S_{1z}^{\rm can})^2=$const.) \lq\lq canonical" spin vector, namely $\{S_{1x}^{\rm can},S_{1y}^{\rm can}\}=S_{1z}^{\rm can}$, etc., ensure that the (linear-in-spin) spin-orbit interaction $H_{\rm SO}=\bfOmega_1^{\rm SO}\cdot {\mathbf S}_1^{\rm can}+\bfOmega_2^{\rm SO}\cdot {\mathbf S}_2^{\rm can}$ implies spin-evolution equations of the form $\frac{d {\mathbf S}_1^{\rm can}}{dt}=\bfOmega_1^{\rm SO} \times {\mathbf S}_1^{\rm can}$, so that the coefficient $\bfOmega_1^{\rm SO}$ entering the spin-orbit Hamiltonian is simply equal to the vectorial precession frequency of the canonical spin vector ${\mathbf S}_1^{\rm can}$. However, one must  note that the components $S_{1x}^{\rm can},S_{1y}^{\rm can}, S_{1z}^{\rm can}$ of the canonical spin vector entering the Hamiltonian have to be defined in a restricted way, compatible with the global $SO(3)$ symmetry of the Hamiltonian dynamics. [The presence of the global $SO(3)$ symmetry  is necessary to ensure, in particular, the conservation of the total angular momentum vector ${\mathbf J}$, having the simple, additive form ${\mathbf J}={\mathbf r}_1\times {\mathbf p}_1+{\mathbf r}_2\times {\mathbf p}_2+{\mathbf S}_1^{\rm can}+{\mathbf S}_2^{\rm can}$.] In Ref. \cite{Damour:2007nc} the compatibility of the definition of ${\mathbf S}_1^{\rm can}$, ${\mathbf S}_2^{\rm can}$
with a global $SO(3)$ symmetry was ensured by relating the components $S_{1a}^{\rm can}$, $(a=1,2,3)$ to the spatial covariant coordinates (in an Arnowitt-Deser-Misner coordinate system $t,x^i$ adapted to the Hamiltonian formulation) of an abstract spin four-vector $S_\mu$, by a specific linear transformation $S_a^{\rm can}=H_a^iS_i$
(where the {\it symmetric} matrix $H_a^i$ depends on $g_{\mu\nu}$ and on the momentum variables). It is easily seen that other $SO(3)$-compatible definitions of $S_a^{\rm can}$ are possible. For instance, one can first associate with the $(3+1)$ [or $(2+1)$] decomposition
\beq
ds^2 =-N^2 dt^2 +\gamma_{ij}(dx^i+N^i dt)(dx^j+N^j dt)
\eeq
the specific moving co-frame $\theta^0=Ndt$, $\theta^a=\theta^a_i (dx^i+N^i dt)$, where $\theta^a_i$ is defined as the {\it symmetric} square-root of $\gamma_{ij}$
(i.e., $\gamma_{ij}=\sum_a \theta^a_i \theta^a_j $ and  $\theta^{a}_i=\theta^i_a$). This specifies a unique, corresponding dual vectorial frame $e_0,e_a$, where $e_0^\mu \partial_\mu=\frac{1}{N}(\partial_t -N^i \partial_i)$ is orthogonal and $e_a=e_a^i\partial_i$ (with $e_a^i\theta^a_j=\delta^i_j$) is {\it tangent}, to the $t=$const. hypersurfaces. 
This frame is $SO(3)$-compatible in the sense that if the matrix $\gamma=(\gamma_{ij})$, with associated symmetric matrix square-root $\theta$ [$\gamma=\theta^{\rm T}\theta$, $\theta^{\rm T}=\theta$] \lq\lq rotates" under an orthogonal transformation $R$, i.e., $\gamma'=R^{\rm T} \gamma R$
(with $R^{\rm T}R=I$), then the unique symmetric matrix square-root $\theta'$ of $\gamma '$ is rotated by the same orthogonal matrix $R$: $\theta'=R^{\rm T} \theta R$.

In terms of this Hamiltonian-adapted (and $SO(3)$-compatible) frame, one can then define $S_{a}^{\rm can}\equiv S^*_\mu e_a^\mu$, where the abstract four-vector $S^{*\mu}$ is tangent to the $t=$const. hypersurface, and is obtained from $S^\mu$ (which is orthogonal to $u^\mu$) by rotating it along the bivector $[e_0 \wedge u]^{\mu\nu}$. [In other words, $S^{*\mu}$ is obtained from $S^\mu$ by a local Lorentz boost associated with the two-plane spanned by $e_0$ and $u$.] Using this specific definition of $S_a^{\rm can}$ (together with the intermediate use of polar coordinates associated with the $(2+1)$-dimensional Cartesian-like coordinates $x^i$; e.g., $x^1=r\cos \phi$, $x^2=r\sin \phi$) we have found that the precession frequency of the Euclidean vector $S_a^{\rm can}$ has as only nonvanishing component $\Omega_z^{\rm SO}=\Omega^{\rm SO}_{xy}$, with
\beq
\label{2.14}
\Omega_z^{\rm SO}=\Omega-K_{\hat 1 \hat 2}=\Omega - |\nabla k|\,,
\eeq
where the contribution $\Omega$ comes from the rotation of the coordinates  $x^1=r\cos (\bar \phi+\Omega t)$, $x^2=r\sin (\bar \phi+\Omega t)$.
Inserting Eqs. (\ref{2.11}), (\ref{2.12}) in Eq. (\ref{2.14}) finally yields an explicit expression for the spin-orbit coupling term $\bfOmega_1^{\rm SO}\cdot {\mathbf S}_1^{\rm can}$ in terms of the covariant components of the ($\theta=\pi/2$)-reduced metric $ds^2$, expressed in polar coordinates $t,r,\phi$.

\section{Spin precession in perturbed Schwarzschild spacetimes}

The results (\ref{2.11}), (\ref{2.14}) of the previous section for the spin-orbit coupling of ${\mathbf S}_1$ are formally valid for general, aligned-spin binaries, with arbitrary mass ratio $m_1/m_2$, and arbitrary ${\mathbf S}_2$, but to only linear order in ${\mathbf S}_1$. Let us now consider the case where $m_1\ll m_2$ and 
${\mathbf S}_2=0$. In that case one is dealing with linear perturbations $h_{\mu\nu}(x^\lambda)$ of a Schwarzschild background of mass $m_2$ by a small mass $m_1$ moving on a circular orbit of radius $r_0$; see Eq. (\ref{1.9}). Because of the Killing symmetry $k=\partial_t +\Omega \partial_\phi$, the metric perturbation depend only on $\bar \phi=\phi-\Omega t$, $r$ and $\theta$, $h_{\mu\nu}(\bar \phi , r, \theta)$. 

The four-velocity of $m_1$ (normalized with respect to the regularized metric $g^R_{\mu\nu}=g^{(0)}_{\mu\nu}+q h_{\mu\nu}^{\rm R}+O(q^2)$, where $q\equiv m_1/m_2\ll 1$) can be written as
\beq
\label{3.1}
u^\mu=\frac{k^\mu}{|k|}\equiv \Gamma k^\mu\,,\qquad \Gamma \equiv \frac{1}{|k|}\,,
\eeq
where (to linear order in $q$)
\begin{eqnarray}
\label{3.2}
|k| &=& \sqrt{[-g_{\mu\nu}^{\rm R} k^\mu k^\nu]_1}=\sqrt{1-\frac{2m_2}{r_0}-\Omega^2 r_0^2-qh_{kk}}\nonumber\\
&=& \sqrt{1-\frac{2m_2}{r_0}-\Omega^2 r_0^2}\left(1-\frac{1}{2}q \frac{h_{kk}}{1-\frac{2m_2}{r_0}-\Omega^2 r_0^2} \right)\nonumber\\
\end{eqnarray}
with $h_{kk}=[h_{\mu\nu}^{\rm R}(x)k^\mu k^\nu]_1$, and
\beq
\label{3.3}
\Gamma =\frac{1}{\sqrt{1-\frac{2m_2}{r_0}-\Omega^2 r_0^2}}\left(1+\frac{1}{2}q \frac{h_{kk}}{1-\frac{2m_2}{r_0}-\Omega^2 r_0^2} \right)\,.
\eeq
The conditions $\partial_\mu g^R_{kk}=0$ for geodesic motion imply
\begin{eqnarray}
\label{3.4}
\Omega &=& \sqrt{\frac{m_2}{r_0^3}}\left(1-q \frac{r_0}{4m_2}[\partial_r h_{kk}^{\rm R}]_1  \right)\,,\\
\label{3.5}
[\partial_{\bar \phi} h_{kk}^{\rm R}]_1&=&0\,.
\end{eqnarray}
Eq. (\ref{3.4}) \cite{Detweiler:2008ft} allows one to trade the gauge-dependent radius $r_0$ for the gauge-invariant dimensionless frequency parameter
\beq
\label{3.6}
y=(m_2\Omega)^{2/3}\,.
\eeq
Namely, we have
\begin{eqnarray}
\label{3.7}
r_0 &=& \frac{m_2}{y}-q \frac{m_2^2}{6y^3}[\partial_r h_{kk}^{\rm R}]_1\,,\nonumber\\
\label{3.8}
\frac{m_2}{r_0}&=&y \left(1+q \frac{m_2}{6y^2}[\partial_r h_{kk}^{\rm R}]_1 \right)\,.
\end{eqnarray}
By inserting Eq. (\ref{1.9}) into Eq. (\ref{2.11}), and using Eq. (\ref{3.7}) to trade $r_0$ for $y$, we get the following explicit expression
of $|\nabla k|$:
\beq
\label{3.9}
m_2 |\nabla k|= y^{3/2}\sqrt{1-3y}\, (1+q\, \delta(y)+O(q^2))\,,
\eeq
where
\begin{eqnarray}
\label{3.10}
\delta (y) &=&  
 -\frac12 (1-2y)h_{rr}-\frac{y^2(1-y)}{2m_2^2 (1-2y)}h_{\phi\phi}\nonumber\\
&& -\frac{y^{3/2}}{m_2(1-2y)}h_{t\phi}-\frac{y}{2(1-2y)(1-3y)}h_{kk}\nonumber\\
&& -\frac{1}{2\sqrt{y}}(\partial_{\bar \phi} h_{rk}-\partial_r h_{\phi k})  \,.
\end{eqnarray}
Here it is understood that all quantities are regularized and evaluated for $\theta=\pi/2$.
The zeroth order term, equivalent (in view of $m_2\Omega=y^{3/2}$) to
\beq
\label{3.11}
|\nabla k|^{(0)}=\Omega \sqrt{1-3y}\,,
\eeq 
is the well-known result for gyroscopic precession (with respect to a rotating, polar-coordinate frame) in a Schwarzschild background
\cite{Straumann}.

The quantity $\delta(y)$ measures the fractional 1SF correction to $|\nabla k|$. Note that the radial derivative $\partial_r h_{kk}$ of $h_{kk}$ 
does not appear in $\delta (y)$. Indeed, the contribution coming from the replacement (\ref{3.7}) has cancelled a similar term that was present in the 1SF expansion of the exact formula (\ref{2.11}). We have checked that the expression (\ref{3.10}) is equivalent to the results given in Ref. \cite{Dolan:2013roa}.
We have also explicitly checked that $\delta(y)$ is gauge-invariant. More precisely, inserting in (\ref{3.10}) the effects of an infinitesimal gauge variation \cite{Detweiler:2008ft}, i.e.
$\Delta h_{\mu\nu}=-\pounds_\xi g_{\mu\nu}=-\nabla_\mu \xi_\nu-\nabla_\nu \xi_\mu$, leads to: $\Delta h_{t\theta}=\Delta h_{r\theta}=\Delta h_{\theta\phi}=0$ and
\begin{eqnarray}
\label{Delta_metric}
\Delta h_{tt}&=&  f'\xi^r +2\Omega  \partial_{\bar \phi}\xi_t \nonumber\\  
\Delta h_{tr}&=& -\frac{1}{f}\left[-\Omega \partial_{\bar \phi} \xi^r +f \partial_r \xi_t -f'\xi_t\right]\nonumber\\ 
\Delta h_{t\phi}&=& \Omega \partial_{\bar \phi} \xi_\phi -\partial_{\bar \phi} \xi_t  \nonumber\\
\Delta h_{rr}&=& \frac{f'}{f^2} \xi^r -\frac{2}{f} \partial_r \xi^r    \nonumber\\
\Delta h_{r\phi}&=& -  \partial_r \xi_\phi +\frac{2}{r} \xi_\phi -\frac{1}{f} \partial_{\bar \phi}\xi^r \nonumber\\
\Delta h_{\theta\theta}&=& -2 r\xi^r \nonumber\\
\Delta h_{\phi\phi}&=& -2 r\xi^r  -2\partial_{\bar \phi} \xi_\phi\,,
\end{eqnarray} 
where $f(r)=1-2m_2/r$.  
We found $\Delta \delta =0$, under the assumption that $\xi_\mu=\xi_\mu(\bar \phi, r, \theta)$.
Evidently, this first order gauge-invariance was expected as the exact function $|\nabla k|(\Omega)$ is defined in a gauge-invariant way.

Note that the quantity of most physical interest is the spin-orbit coupling (\ref{2.14}), which reads, to first order in $q$
\beq
m_2\Omega_1^{\rm SO}(y)= y^{3/2}-y^{3/2}\sqrt{1-3y}(1+q\, \delta(y))\,.
\eeq
Here the unperturbed value of the function $\Omega_1^{\rm SO}(y)$ is given by
\beq
m_2\Omega_1^{{\rm SO }(0)}(y)= y^{3/2}-y^{3/2}\sqrt{1-3y}\,,
\eeq
so that the 1SF correction to $\Omega_1^{\rm SO}(y)=\Omega_1^{{\rm SO }(0)}(y)+q\, \Omega_1^{\rm SO (1SF)}(y)+O(q^2)$ is related to the notation $\delta(y)$ used above via
\beq
m_2\Omega_1^{\rm SO (1SF)}(y)= -\, y^{3/2}\sqrt{1-3y}\, \delta(y)\,.
\eeq

\section{Analytic computation of $\delta(y)$ in the Regge-Wheeler-Zerilli formalism}

As our aim is to compute the gauge-invariant quantity $\delta(y)$, Eq. (\ref{3.10}), we should be able to compute it in any (asymptotically flat) gauge.

Similarly to our recent works \cite{Bini:2013zaa,Bini:2013rfa,Bini:2014nfa} where we analytically computed the zero-derivative-level, gauge-invariant quantity $h_{kk}(y)$, we shall use the Regge-Wheeler-Zerilli (RWZ) formalism and work in Regge-Wheeler gauge. We refer to Refs. \cite{Bini:2013rfa,Bini:2014nfa} for technical details, and elaborate here only the 
new issues that we had to face.

As the RWZ perturbation $h_{\mu\nu}$ is decomposed, for each multiple order $(l,m)$, in even-parity and odd-parity pieces, we can correspondingly decompose the multipolar piece $\delta^{(l,m)}(y)$ of $\delta(y)$, Eq. (\ref{3.10}), (which is linear in $h_{\mu\nu}$ and its first derivatives) as
\beq
\label{4.1}
\delta^{(l,m)}(y)=\delta_{\rm (even)}^{(l,m)}(y)+\delta^{(l,m)}_{\rm (odd)}(y)\,.
\eeq
In terms of the usual RWZ notation, and suppressing the $l,m$ indices, we find  the following expressions for the \lq\lq bare" (i.e., unregularized) values of 
$\delta_{\rm (even/odd)}(y)$ 
\begin{eqnarray}
\label{bare_delta}
\delta^{\rm bare}_{\rm (odd)}&=& \left[-\frac{\Omega r}{(r-3m_2)}h_0  +\frac{1}{2\Omega r}h_0'\right. \nonumber\\
&& \left. -\frac{im}{2r} h_1  \right]\frac{dY}{d\theta} \left(\frac{\pi}{2},0\right)\nonumber\\
\delta^{\rm bare}_{\rm (even)}&=& \left[-\frac{m_2}{2(r-3m_2)}H_0  -\frac{im}{2\Omega r} H_1-\frac12  H_2\right.\nonumber\\
&& \left. +\frac{(r-4m_2)}{2(r-3m_2)}K+\frac12 r K' \right]Y\left(\frac{\pi}{2},0\right)\,.   
\end{eqnarray}

As in our previous work, we used results of the RWZ literature \cite{Sago:2002fe,Nakano:2003he,Hikida:2004jw,Chandrasekhar:1975} to express the metric perturbation $h_{\mu\nu}^{(l,m)}$, with frequency $\omega=m\Omega$, in terms of radial functions $R_{lm\omega}^{\rm (even/odd)}(r)$ satisfying the simple Regge-Wheeler equation
\beq
{\mathcal L}_{\rm (RW)}^{(r)}[R_{lm\omega}^{\rm (even/odd)}(r)]=S_{lm\omega}^{\rm (even/odd)}(r)
\eeq
where
\beq
{\mathcal L}_{\rm (RW)}^{(r)}=\frac{d^2}{dr_*^2} +\omega^2 -\left(1-\frac{2m_2}{r}  \right)\left(\frac{l(l+1)}{r^2}-\frac{6m_2}{r^3}  \right)\,,
\eeq
with $dr^*=dr/f(r)$ and $f(r)=1-\frac{2m_2}{r}$.
Here, the odd-parity source is a combination of $\delta(r-r_0)$ and $\delta'(r-r_0)$, while the even-parity one is a combination of $\delta(r-r_0)$, $\delta'(r-r_0)$ and 
$\delta ''(r-r_0)$. The inhomogeneous radial equations are solved, as usual, by means of a suitable radial Green function $G(r,r')$:
\beq
\label{eq33}
R_{lm\omega}^{\rm (even/odd)}(r)=\int dr' G(r,r') f(r')^{-1}S_{lm\omega}^{\rm (even/odd)}(r')\,.
\eeq
Note that the distributional nature of the radial source functions (notably $S_{lm\omega}^{\rm (even)}(r)\ni \delta ''(r-r_0)$)   implies the presence of a correspondingly discontinuous (or even distributional) terms in the solution $R_{lm\omega}^{\rm (even/odd)}(r)$ as $r$ crosses the radial location 
$r_0$ of the source (i.e., the orbiting small mass $m_1$).

In the case of the less singular $h_{kk}^{(l,m)}(x^\lambda)$ we had found that, in Regge-Wheeler gauge, $h_{kk}^{(l,m)}(r)$ was continuous across $r_0$, i.e. that the two limits $r\to r_0^+$ and $r\to r_0^-$ (keeping $\theta$ and $\phi$ to the values corresponding to the considered instantaneous position of particle 1, say
$t=t_1$, $\theta=\pi/2$ and $\phi=\Omega t_1$) did coincide.
As we are now considering a more singular quantity, involving first derivatives of $h_{\mu\nu}(x^\lambda)$, we cannot expect to have such a simple behaviour.
However, similarly to the evaluations of the (gauge-dependent) gravitational self-force \cite{Barack:2009ux,Barack:1999wf,Barack:2001gx} $\propto \Gamma^\lambda{}_{\mu\nu}u^\mu u^\nu\sim \partial_\lambda h_{\mu\nu}$, we expect that taking the average over the two radial limits ($r_0^+$ and $r_0^-$) will eliminate the worst singularity (due to $\partial_\mu \rho^{-1}$ where $\rho $ denotes the distance between the field point and the world line) and will leave only a milder singularity, corresponding to a milder growth with the multipolar order $l$. This is indeed what we found: while the \lq\lq jump" $\delta^+_{lm}-\delta^-_{lm}$ (where $\delta^\pm_{lm}=\lim_{r\to r_0^\pm}\delta _{lm}(r)$) grows (after summing over the \lq\lq magnetic" number $m$) proportionally to $2l+1$, the averaged $\delta$ (where we sum over the even and odd contributions, as well as over $m$)
\beq
\langle \delta^l \rangle =\sum_{m=-l}^l
\frac12 (\delta^+_{lm}+\delta^-_{lm})
\eeq
was found to have a limit, say $\langle \delta^\infty \rangle$, as $l\to \infty$.

In keeping with standard GSF results on mode regularization \cite{Barack:1999wf,Barack:2001gx}, we can then compute the regularized value of $\delta$ as
\beq
\label{4.6}
\delta^{\rm R}(y)=\sum_{l=0}^\infty [\langle \delta^l (y)\rangle - \langle \delta^\infty \rangle]\,.
\eeq
Let us now give some details of our analytical evaluation of $\langle \delta^l (y)\rangle$, $\langle \delta^\infty \rangle$ and $\delta^{\rm R}(y)$.
Our starting point is the expression of the radial Green's function entering the solutions:
\begin{eqnarray}
\label{eq31}
G(r,r')&=&\frac{1}{W}\Bigl[X_{\rm (in)}(r)X_{\rm (up)}(r')H(r'-r) \nonumber \\
&+&X_{\rm (in)}(r')X_{\rm (up)}(r)H(r-r')  \Bigl]\,, 
\end{eqnarray}
where $X_{\rm (in/up)}(r)$ are solutions of the homogeneous RW equation, $H(x)$ is the Heaviside step function and where $W$ denotes the (constant) Wronskian
\begin{eqnarray}
\label{eq32}
W&=&\left(1-\frac{2m_2}{r}  \right)\biggl[X_{\rm (in)}(r)\frac{d}{dr }X_{\rm (up)}(r )\nonumber \\
&-&\frac{d}{dr}X_{\rm (in)}(r)X_{\rm (up)}(r) \biggl]={\rm constant} \,.
\end{eqnarray}
The physically relevant  Green's function here is the   retarded  one, corresponding to an $X_{({\rm in})}$  
incoming from $r = +\infty$ (and purely ingoing on the horizon), and to an $X_{({\rm up})}$ 
upgoing from the horizon (and purely outgoing at infinity).
With this specification the solutions,  Eq. (\ref{eq33}), of the {\it inhomogeneous} even-parity and odd-parity Regge-Wheeler-Zerilli equations
are unique.

The second step of our strategy consists  in computing  the explicit expressions of $\delta^{\pm {\rm even/odd}}_{lm}$ in terms of the radial solutions $X_{({\rm in})}$ and $X_{({\rm up})}$ of the Regge-Wheeler homogeneous equation, Eqs. (\ref{bare_delta}).
For instance, for the $r=r_0^-$ limit, we found (with ${}'\equiv d/dr$)
\begin{widetext}
\beq
\label{5.4notes}
\delta^{-\rm (odd)}_{lm}=  \frac{8\pi }{(l+2) (l-1) (l+1) l} \frac{1}{ (1-2y)  (1-3y)^{3/2}} \frac{ \left(A_{\rm (in)}^{\rm (odd)} X_{\rm (in)} +B_{\rm (in)}^{\rm (odd)} m_2 X_{\rm (in)}'\right)\left(y X_{\rm (up)} +m_2 X_{\rm (up)}' \right)}{(X_{\rm (in)} m_2 X_{\rm (up)}' -X_{\rm (up)} m_2 X_{\rm (in)}' )  }\left|\frac{dY_{lm}}{d\theta}\right|^2 
\eeq
where [with the notation $L=l(l+1)$, $\lambda=(l-1)(l+2)/2$, $\Lambda=\lambda (\lambda +1)$]
\begin{eqnarray}
A_{\rm (in)}^{\rm (odd)} 
&=& y(32 y^3-6 y^2 L -28 y^2-6   m^2 y^2+5 y L+6 y +2 y m^2 -L) \nonumber\\
B_{\rm (in)}^{\rm (odd)}  
&= &2 \left(1-2y\right)^2 \left(1-4y\right)\,,
\end{eqnarray}
and
\beq
\label{5.6notes}
\delta^{-\rm (even)}_{lm}=    \frac{4\pi }{\Lambda (\Lambda^2 +9m^2y^3)} \frac{\left(A_{\rm (in)}^{\rm (even)} X_{\rm (in)} +B_{\rm (in)}^{\rm (even)}m_2 X_{\rm (in)}'\right)\left(A_{\rm (up)}^{\rm (even)} X_{\rm (up)} +B_{\rm (up)}^{\rm (even)}m_2 X_{\rm (up)}' \right)}{(1-2y)^2(1-3y)^{3/2}(X_{\rm (in)}m_2 X_{\rm (up)}' -X_{\rm (up)}m_2 X_{\rm (in)}' )}
\left| Y_{lm}\right|^2 
\eeq
where
\begin{eqnarray}
A_{\rm (in)}^{\rm (even)}&=& y (3 m^2 y-12 y^3\lambda^2-3 m^4 y^2+m^2\lambda^2+11 y\lambda^2-6 m^2 y\lambda^2+26 y^2\lambda-3 m^2 y\lambda\nonumber\\
&& -10 y^2\lambda^3-8 y^2\lambda^2-72 y^3\lambda-\lambda^3-2\lambda^2+m^2\lambda-60 y^3+y \lambda+7 y\lambda^3+9 m^4 y^3+24 y^2\nonumber\\
&& +9 m^2 y^3\lambda-48 m^2 y^4+48 y^4+48 y^4\lambda-4 y^2 m^2\lambda+8 y^2 m^2\lambda^2-21 m^2 y^2+51 m^2 y^3-\lambda-3 y)\nonumber\\
B_{\rm (in)}^{\rm (even)}&=&  3m^2y+4y^3\lambda^2+m^2\lambda^2+11y\lambda^2-5m^2y\lambda^2+14y^2\lambda-5m^2y\lambda-6y^2\lambda^3-16y^2\lambda^2\nonumber\\
&&-56y^3\lambda-\lambda^3-2\lambda^2+m^2\lambda-60y^3+3y\lambda+5y\lambda^3+24y^2-48m^2y^4+48y^4+48y^4\lambda+6y^2m^2\lambda\nonumber\\
&&+6y^2m^2\lambda^2-24m^2y^2+60m^2y^3-\lambda-3y \nonumber\\
A_{\rm (up)}^{\rm (even)}&=&  y(-6y^3+2 \lambda^2-y \lambda^3-4y \lambda-y m^2\lambda-6y^3\lambda-5y \lambda^2+9y^2\lambda+\lambda^3 -y m^2\lambda^2\nonumber\\
&&-3m^2y^2+6m^2y^3+3y^2+ \lambda+6y^2\lambda^2) \nonumber\\
B_{\rm (up)}^{\rm (even)}&=& \lambda^2-6y^3+3y^2+7y^2\lambda+4y^2\lambda^2+\lambda-3m^2y^2+6m^2y^3-4y\lambda-6y^3\lambda-4y\lambda^2  
\,.
\end{eqnarray}
\end{widetext}
The corresponding results for the outer limit $r\to r_0^+$ are given by similar expressions.

The third step of the  method consists of breaking up the analytical evaluation of $\langle\delta^l \rangle$ into three parts: (i) the non-radiative multipoles, $0\le l \le 1$; (ii) several of the low radiative multipoles, $2 \le l \le l_{\rm max}$; and (iii) the generic, higher radiative multipoles, $l\ge l_{\rm max}+1$. For the part (i),  $0\le l \le 1$, we use the analytic results of Zerilli \cite{Zerilli:1970se} (see also \cite{Detweiler:2003ci}), except for the fact that they must be formulated
in an asymptotically flat gauge. Explicitly, we have the following low-multipole metric perturbations.
For $l=0$
\begin{eqnarray}
h_{tt}&=& \frac{2 \tilde E_1}{r_0}\frac{1-\frac{2m_2}{r}}{1-\frac{2m_2}{r_0}}H(r_0-r)+\frac{2  \tilde E_1}{r}H(r-r_0)\nonumber\\
h_{rr}&=&  \frac{2  \tilde E_1}{r\left( 1-\frac{2m_2}{r} \right)^2}H(r-r_0)\,,
\end{eqnarray}
with
\beq
\tilde E_1= \frac{1-\frac{2m_2}{r_0}}{\sqrt{1-\frac{3m_2}{r_0}}}+O(q)= \frac{1-2y}{\sqrt{1-3y}}+O(q)\,.
\eeq
For $l=1$ (odd) 
\begin{eqnarray}
h_{t\phi}&=& -2  \tilde L_1 \sin^2\theta \left[\frac{r^2}{r_0^3}H(r_0-r) +\frac1r  H(r-r_0)\right] \,,
\end{eqnarray}
with
\beq
\tilde L_1 =  \sqrt{ \frac{r_0}{m_2(1-\frac{3m_2}{r_0})}}+O(q)= \frac{1}{\sqrt{y(1-3y)}}+O(q)\,.
\eeq
For $l=1$ (even) 
\begin{eqnarray}
h_{tt}&=& -2  \tilde E_1 \frac{r_0-2m_2}{r(r-2m_2)}\left(1-\frac{r^3\Omega^2}{m_2}  \right)\sin \theta \cos \bar \phi H(r_0-r) \nonumber\\
h_{tr}&=& 6  \tilde E_1 \Omega \frac{r(r_0-2m_2)}{(r-2m_2)^2} \sin \theta \sin \bar \phi H(r_0-r)\nonumber\\
h_{rr}&=& -6  \tilde E_1  \frac{r(r_0-2m_2)}{(r-2m_2)^3} \sin \theta \cos \bar \phi H(r_0-r)\,.
\end{eqnarray}
The corresponding contributions to $\langle \delta^l \rangle$ (without the subtraction $\langle \delta^\infty \rangle$ that we shall discuss below) are obtained by inserting the above results in the general Eq. (\ref{3.10}). In these particular cases, the right hand side of (\ref{3.10}) simplifies to
\begin{eqnarray}
\delta^{l=0}&=&   -\frac12 (1-2y)h_{rr}-\frac{y}{2(1-2y)(1-3y)}h_{tt} \nonumber\\
\delta^{l=1\, {\rm (odd)}}&=& -\frac{y^{3/2}}{1-2y}\frac{1}{m_2}h_{t\phi}-\frac{y^{5/2}}{(1-2y)(1-3y)}\frac{1}{m_2} h_{t\phi}\nonumber\\
&& +\frac{1}{2\sqrt{y}}\partial_{r}h_{t\phi}\nonumber\\
\delta^{l=1\, {\rm (even)}}&=&  -\frac12 (1-2y)h_{rr}-\frac{1}{2\sqrt{y}}m_2 \partial_{\bar \phi}h_{rt}\,.
\end{eqnarray}
After inserting the corresponding expressions of the metric perturbations and taking the average between the outer and the inner radial limits $r\to r_0^\pm$, the corresponding results for $\langle \delta^l \rangle$ read
\begin{eqnarray}
\langle \delta^{l=0} \rangle &=& -\frac{y(1-y)}{2(1-3y)^{3/2}}\,,\nonumber\\
\langle \delta^{l=1\, {\rm odd}} \rangle &=& -\frac{y(1-7y)}{2(1-3y)^{3/2}}\,, \nonumber\\
\langle \delta^{l=1\, {\rm even}} \rangle &=& 0 \,.
\end{eqnarray}
Note that the even $l=1$ multipole gives no contribution, and that the sum of all the (unregularized) low multipole contributions reads
\beq
\label{td4.22}
\langle \delta^{0\le l \le 1} \rangle= -\frac{y(1-4y)}{(1-3y)^{3/2}}\,.
\eeq
For the intermediate radiative multipoles $2\le l \le l_{\rm max}$ we used the results of Mano, Suzuki and Tagasugi (MST) \cite{Mano:1996vt,Mano:1996mf,Mano:1996gn}.
MST gave analytic expressions for $X_{\rm (in)}$ and $X_{\rm (up)}$ in the form of series of hypergeometric functions.
We refer to our previous papers \cite{Bini:2013zaa,Bini:2013rfa,Bini:2014nfa} for details on the explicit implementation of the MST expansions. As discussed there, one must choose the value of $l_{\rm max}$ according to the PN accuracy that one is aiming at. In the present work, we aim at getting $\delta(y)$ up to the eight-and-a-half PN accuracy \footnote{This counting amounts to considering that the leading order PN contribution to $\Omega_1^{\rm SO}/\Omega=1-\sqrt{1-3y}(1+q\, \delta(y))\simeq \frac32 y -q\delta(y)$ is of 1PN order, which is the usual PN counting for spin-orbit effects.}, i.e., up to $O(y^{17/2})$ included. To reach this accuracy,
 we used the MST expansions up to the multipole order $l_{\rm max}=5$ (see below).
For generic, higher multipoles $l\ge l_{\rm max}+1$ we used the PN-expanded solutions for $X_{\rm (in)}$ and $X_{\rm (up)}$, following the method of our previous works \cite{Bini:2013zaa,Bini:2013rfa,Bini:2014nfa}. 

Before giving our final results, let us illustrate some of the intermediate results we got by our analytic approach.
Let us first point out that, starting from the explicit expressions, Eqs. (\ref{5.4notes})-(\ref{5.6notes}), of the $\delta_{lm}^{\pm {\rm (even/odd)}}$ and using the equations satisfied by $X_{\rm (in)}$ and $X_{\rm (up)}$, we could derive exact expressions for the jumps $[\delta_{lm}]=[\delta_{lm}^{+}-\delta_{lm}^{-}]$ of $\delta_{lm}(r)$ across $r_0$. 

Explicitly, the (half) odd and even jumps are given by
\begin{widetext}
\begin{eqnarray}
\frac12 [\delta_{lm}^{ {\rm (odd)}}]&=&  -\frac{4\pi   y}{(1-2y)\sqrt{1-3y}}  \frac{[2-l(l+1)](1-2y)+2m^2y }{l(l+1)(l-1)(l+2)}\left|\frac{dY_{lm}}{d\theta}\left( \frac{\pi}{2},0 \right)\right|^2 \nonumber\\
\frac12 [\delta_{lm}^{ {\rm (even)}}]&=& \frac{2\pi  y}{ (1-2 y)\sqrt{1-3y} }  \frac{(l(l+1)-2 m^2)[(y-1)l(l+1)+2-4 y+2 m^2 y] }{(l+1) l (l-1) (l+2)} \left|Y_{lm}\left( \frac{\pi}{2},0 \right) \right|^2 \,.
\end{eqnarray}
\end{widetext}
Summing over $m$ (using the same summation formulas we used in previous works \cite{Nakano:2003he}) thus yields
\begin{eqnarray}
\frac12 [\delta_{l}^{ {\rm (odd)}}]&=&   -  (2l+1)\frac{y(5y-2)}{4(1-2y)\sqrt{1-3 y}}\,, \nonumber\\
\frac12 [\delta_{l}^{ {\rm (even)}}]&=&  - (2l+1)\frac{ y^2}{4(1-2y)\sqrt{1-3y}} \,,\nonumber\\
\frac12 [\delta_l]&\equiv & \frac12 [\delta_{l}^{ {\rm (odd)}}+\delta_{l}^{ {\rm (even)}}]\nonumber\\
&=&(2l+1)\frac{y\sqrt{1-3y}}{2(1-2y)}\,.
\end{eqnarray}
As announced above, these jumps are proportional to $(2l+1)$ which is compatible with the GSF-like behavior $\pm A(l+1/2)+B+O(1/l^2)$ \cite{Barack:1999wf,Barack:2001gx} expected for the large $l$ multipolar decomposition of a quantity containing  first derivatives of $h_{\mu\nu}$.

As already mentioned, it is then convenient to eliminate the leading order term $\pm A(l+1/2)$ by working with the radial average $\langle \delta_l \rangle \equiv \frac12 (\delta_l^++\delta_l^-)$.
No closed-form analytic expression can be given for $\langle \delta_l \rangle \equiv \sum_m  \langle \delta_{lm} \rangle$. We can, however, either compute its PN expansion in powers of $y$ for any integer value of $l$ by using MST-like expansions or, for a generic, unspecified value of $l$ (which can even be considered as taking any real or complex value) we can compute its PN expanded expression by using the PN-expanded solutions of $X_{\rm (in)}$ and $X_{\rm (up)}$ \cite{Bini:2013zaa,Bini:2013rfa,Bini:2014nfa}.
Let us quote, for illustration, the 3PN (i.e., $O(y^3)$ accurate) truncation of our high-PN-order results.
We found (after summing over $m$)
\begin{widetext}
\begin{eqnarray}
\delta_{l \rm (odd)}^{-}&=& - (l+1) y
-\frac{  (l+1) (16 l^2-9 l-12) y^2}{4(-1+2 l) (2 l+3)}\nonumber\\
&& -\frac{ (208 l^7+179 l^6-1647 l^5-1346 l^4+2510 l^3+1767 l^2-471 l+2160) y^3}{8l (-1+2 l) (2 l+3) (2 l-3) (2 l+5) (l+1)} +O(y^4)\,,\nonumber\\
\delta_{l\rm (odd)}^{+}&=& \delta_{l\rm (odd)}^{-}|_{_{l\to -l-1}}\,,\nonumber\\
\delta_{l\rm (even)}^{-}&=&  \frac{(8 l^3-l^2-15 l-12)y^2}{4(-1+2 l) (2 l+3)}  \nonumber\\
&& +\frac{  (224 l^7+298 l^6-1458 l^5-1219 l^4+2638 l^3+1161 l^2-1164 l-2160)y^3}{8l(-1+2l) (2 l+3) (2 l-3) (2 l+5) (l+1)}  +O(y^4)\,, \nonumber\\  
\delta_{l\rm (even)}^{+}&=& \delta_{l \rm (even)}^{-}|_{_{l=-l-1}}\,.
\end{eqnarray}  
\end{widetext}
Note the simple \lq\lq symmetry" between $\delta_l^{-}$ and $\delta_l^{+}$: the outer limit $\delta_l^+$ is obtained from the inner one $\delta_l^{-}$ simply by replacing $l$ by $-l-1$. In terms of $\bar l\equiv l+1/2$ this replacement reads $\bar l\to -\bar l$. It ensures that $\langle \delta_l \rangle$ will end up being an even function of $\bar l$, expressible in terms of ${\bar l}^2$ only. The explicit 3PN-accurate expressions of $\langle \delta_l^{\rm (even)} \rangle$ and  $\langle \delta_l^{\rm (odd)} \rangle$ from the results above read
\begin{widetext}
\begin{eqnarray}
\label{delta_l_odd_and_even}
\langle \delta_l^{\rm (odd)}(y) \rangle&=&    -\frac12 y
+\frac{(17 l^2+17 l+6)  y^2}{4(-1+2 l)(2 l+3)} \nonumber\\
&& +\frac{3 (366 l^6+1098 l^5-316 l^4-2462 l^3-905 l^2+509 l-1440)y^3}{16l (-1+2 l) (2 l+3) (2 l-3) (2 l+5) (l+1)} +O(y^4)
\nonumber\\{}
\langle \delta_l^{\rm (even)}(y) \rangle&=&   -\frac{(13 l^2+13 l+9)y^2}{4(-1+2 l) (2 l+3)} -\frac{3 (162 l^6-247 l^4-1304 l^3-240 l^2+493 l+486 l^5+720)y^3}{8l (-1+2 l) (2 l+3) (2 l-3) (2 l+5) (l+1)} +O(y^4)\,.
\end{eqnarray}
\end{widetext}
The total result $\langle \delta_l\rangle=\langle \delta_l^{\rm (even)} \rangle+\langle \delta_l^{\rm (odd)} \rangle$ reads
\begin{eqnarray}
\label{sum_delta_l_odd_and_even}
&& \langle \delta_l (y) \rangle= -\frac12 y+\frac14  y^2+ \frac{3y^3}{16 l (l+1)}\times \nonumber\\
&& \frac{(42 l^6+126 l^5+178 l^4+146 l^3-425 l^2-477 l-2880)}{ (2 l-1) (2 l+3) (2 l-3) (5+2 l)}  \nonumber\\
&&\quad  +O(y^4)\,.
\end{eqnarray}

At this stage, let us point out one convenient technical feature of our approach. As exemplified on Eqs. (\ref{delta_l_odd_and_even}) and (\ref{sum_delta_l_odd_and_even}), our PN-expanded, generic-$l$ approach yields explicit analytic expressions for the $l$-dependence of $\langle \delta_l(y) \rangle$. In particular, we see on Eq. (\ref{sum_delta_l_odd_and_even}) that $\langle \delta_l(y) \rangle$ admits a finite limit as $l\to \infty$, which can be easily read off from Eq. (\ref{sum_delta_l_odd_and_even}), viz
\begin{eqnarray}
\label{delta_l_infty}
\langle \delta^\infty (y) \rangle&=&\lim_{l\to \infty}\langle \delta_l (y)\rangle \nonumber\\
&=& -\frac{1}{2}y+\frac14 y^2+\frac{63}{128}y^3+O(y^4)\,. 
\end{eqnarray}
We can therefore {\it regularize} $\langle \delta^\infty (y) \rangle$, according to Eq. (\ref{4.6}), without having to derive, in advance, the analytic expression of the \lq\lq B term" in the large-$l$ expansion
\beq
\delta_l^\pm (y)= \pm \left(l+\frac12 \right) A(y)+B(y)+O\left(\frac{1}{l^2} \right)
\eeq
of $\delta^l$, i.e., the constant term in the large-$l$ expansion of the radial average
\beq
\langle \delta_l (y) \rangle =B(y)+O\left(\frac{1}{l^2} \right)\,.
\eeq
From our high-PN-order results, we have so determined the PN expansion of $B(y)=\langle \delta^\infty (y) \rangle $ to the 8.5PN accuracy, namely
\begin{eqnarray}
\label{B_PN}
B^{\rm PN}(y)&=&\langle \delta^\infty (y) \rangle^{PN}\nonumber\\
&=&  -\frac{1}{2}y+\frac14 y^2+\frac{63}{128}y^3+\frac{995}{1024}y^4\nonumber\\
&& +\frac{63223}{32768}y^5+\frac{126849}{32768}y^6+\frac{1909935}{262144} y^7\nonumber\\
&& +\frac{709638057}{67108864}y^8+O(y^9)\,.
\end{eqnarray}
Continuing the low-PN-order illustration of our method, the $B$-subtracted value of $\langle \delta^\infty (y) \rangle$ obtained from Eq. (\ref{sum_delta_l_odd_and_even}) explicitly reads
\begin{eqnarray}
\label{td4.32}
&&\langle \delta^l (y) \rangle-\langle \delta^\infty (y) \rangle= \nonumber\\
&& \frac{3(3856 l^3+1928 l^4-2833 l^2-4761 l-23040) }{128 l(l+1) (-1+2 l) (2 l+3) (-3+2 l) (5+2 l)}    y^3\nonumber\\
&& 
+O(y^4)\,.
\end{eqnarray}
The coefficient of $y^3$ in this expression is easily seen to be of order $O\left(\frac{1}{l^2} \right)$ as $l\to \infty$.
As mentioned above, this term (as well as the high order ones) is also invariant under the symmetry $l\to -l-1$, so that it could be expressed as a rational function of ${\bar l}^2=(l+1/2)^2$.

Finally, the PN-expanded regularized value $\delta^{\rm R}(y)$ of $\delta$ is obtained from its definition (\ref{4.6}) as
\begin{eqnarray}
\delta^{\rm R}(y)&=& \sum_{l=0}^1(\langle \delta^{l \rm Z} (y) \rangle-B^{\rm PN}(y))\nonumber\\
&& + \sum_{l=2}^5(\langle \delta^{l \rm MST} (y) \rangle-B^{\rm PN}(y))\nonumber\\
&& + \sum_{l=6}^\infty(\langle \delta^{l \rm PN} (y) \rangle-B^{\rm PN}(y))\,.
\end{eqnarray}

Here, the first sum is the contribution of the low multipoles that we derived above from the old results of Zerilli, the second sum comes from using the MST hypergeometric-expansion form of $X_{\rm (in)}$ and $X_{\rm (up)}$, while the third sum comes from the PN-expanded form of $X_{\rm (in)}$ and $X_{\rm (up)}$. As explained above, in all three sums, the subtraction $B^{\rm PN}(y)=\langle \delta^\infty (y) \rangle$ is obtained as a PN-expansion, see Eq. (\ref{B_PN}), from our generic-$l$, PN-expanded $\langle \delta^l (y) \rangle$. From the general-$l$ result  (\ref{td4.32}) we see that all the terms associated with dynamical multipoles $l\ge 2$ will start contributing to $\delta^{\rm R}(y)$ at order $O(y^3)$.
The contributions of order $y$ and $y^2$ can only come from non-dynamical multipoles $l=0,1$. The \lq\lq bare" low multipole contribution (\ref{td4.22}) is equal to $\langle \delta^{0\le l \le 1}\rangle= -y -\frac12 y^2 +O(y^3)$. Subtracting from this   $\sum_{l=0}^1B^{\rm PN}=2B^{\rm PN}=-y+\frac12 y^2+O(y^3)$, we see that $\delta^{\rm R}(y)=-y^2+O(y^3)$.

Our final, 8.5PN-accurate result for $\delta^{\rm R}(y)$ reads
\begin{widetext}
\begin{eqnarray}
\label{4.33new}
\delta^{\rm R}(y)&=& - y^2+\frac{3}{2} y^3+\frac{69}{8} y^4+ (c_5 +c_5^{\ln{}}\ln y)y^5 + (c_6 +c_6^{\ln{}}\ln y)y^6 +\frac{26536}{1575}\pi y^{13/2}\nonumber\\
&& + (c_7 +c_7^{\ln{}}\ln y)y^7 + \frac{670667}{22050} \pi y^{15/2}+(c_8 +c_8^{\ln{}}\ln y+c_8^{\ln^2{}}\ln^2  y)y^8\nonumber\\
&& 
-\frac{3872542979}{13097700}\pi   y^{17/2}+O_{\ln{}}(y^9)\,,
\end{eqnarray}
where
\begin{eqnarray}
c_5&=&\frac{53321}{240} +\frac{496}{15}\ln(2) +16\gamma-\frac{20471}{1024}\pi^2\nonumber\\
c_5^{\ln{}} &=& 8 \nonumber\\
c_6&=& \frac{15462423}{4480} -\frac{357521}{1024}\pi^2+\frac{172}{5}\gamma+ \frac{1436}{105}\ln(2)+\frac{729}{14}\ln(3) \nonumber\\
c_6^{\ln{}} &=&\frac{86}{5} \nonumber\\
c_7&=&-\frac{30832}{105}\gamma-\frac{3344}{21}\ln(2)+\frac{78544852143331829}{5866372512000} -\frac{40581}{140}\ln(3)-\frac{512537515}{393216}\pi^2+\frac{1407987}{524288}\pi^4 \nonumber\\
c_7^{\ln{}} &=&-\frac{15416}{105}\nonumber\\
c_8&=& -\frac{272898799212463348902641}{16830740064378240000} -\frac{1291394011}{3638250}\ln(2)+\frac{96697099}{141750}\gamma+\frac{2364633}{12320}\ln(3)\nonumber\\
&& +\frac{9765625}{28512}\ln(5)-\frac{869696}{1575}\ln(2)\gamma+\frac{1344}{5}\zeta(3)-\frac{58208}{105}\ln(2)^2-\frac{3424}{25}\gamma^2-\frac{63475197004061}{22295347200}\pi^2\nonumber\\
&& +\frac{162286431837}{335544320}\pi^4 \nonumber\\
c_8^{\ln{}} &=&\frac{96697099}{283500} -\frac{434848}{1575} \ln(2)-\frac{3424}{25}\gamma  \nonumber\\
c_8^{\ln^2{}} &=& -\frac{856}{25}\,,
\end{eqnarray}
\end{widetext}
and where $O_{\ln{}} (y^9)$ denotes an error term $O(y^9 (\ln y)^n)$ for some unspecified (integer) power $n$.
The logarithmic terms in $\delta^{\rm R}(y)$ are linked to the back-scattered effect of tails in the near-zone \cite{Blanchet:1987wq,Damour:2009sm,Blanchet:2010zd}. Inserting in the general-$l$ structure of radiation-reaction \cite{Blanchet:1984wm} the (tail-related) hereditary modification of the near-zone reactive multipole moments (which contain an extra factor $GM/c^3$), one finds that the $l$-th mass-type (i.e., even) and spin-type (i.e., odd) multipoles {\it both} contribute to $\Omega_1/\Omega$ (or $\delta^{\rm R}$) at order $O(1/c^{2l+6})$, i.e., at the $(l+3)$-PN level $\sim O(y^{l+3}\ln y)$. [By contrast, the tail-related logarithmic contributions to $h_{kk}(y)$ were of order $O(y^{l+3}\ln y)$ when coming from an even $l$-th reactive multipole and of order  $O(y^{l+4}\ln y)$ when coming from an odd $l$-th reactive multipole; see discussion in Sec. IID of \cite{Bini:2013rfa}.]

At orders $y^5$, $y^6$ and $y^7$, the logarithm of $y$ and the Euler constant $\gamma$ always enter the results through the combination $\ln y +2\gamma$. Similarly, expressing the coefficient of $y^8$ in terms of  $\ln y +2\gamma$ makes the terms of the form $q_1\gamma +q_2\gamma^2$ (with rational coefficients $q_1$ and $q_2$) disappear. As in our previous works 
\cite{Bini:2013zaa,Bini:2013rfa} the structure of the logarithmic terms  depend on the corresponding relevant frequency $\omega_m=m\Omega$ with $|m|\le l$;
e.g., the $l=2$ tail terms contribute $\ln 2$ in $c_5$; the $l\le 3$ tail terms contribute $\ln 3$ and $\ln 2$ in $c_6$; the $l\le 4$ tail terms yield $\ln 4=2\ln 2$, $\ln 3$ and $\ln 2$ in $c_7$; etc. Note also the increase in transcendentality of the coefficients $c_n$ as $n$ increases, with the remarkable first appearance of $\zeta(3)$ in $c_8$. 

\section{Comparison to the numerical GSF results of Dolan et al.}

As already mentioned in the Introduction, Dolan et al. \cite{Dolan:2013roa} have recently numerically evaluated the first-order GSF contribution to the dimensionless ratio between the spin-precession frequency $\Omega_{\rm prec}$ and the orbital frequency $\Omega$:
\beq
\label{5.1}
\psi(y)\equiv \frac{\Omega_{\rm prec}}{\Omega}=1-\frac{|\nabla k|}{\Omega}\,.
\eeq 
To linear order in $q=m_1/m_2$ (keeping $y=(m_2\Omega)^{3/2}$ fixed), $\psi(y)$ reads
\beq
\label{5.2}
\psi (y)=1-\sqrt{1-3y}[1+q\, \delta^{\rm R}(y)+O(q^2)]\,,
\eeq 
so that the 1SF piece $\delta \psi (y)$ in $\psi (y)$ (such that $\psi(y)=\psi_{(0)}(y)+q\, \delta \psi (y)+O(q^2)$) is related to our 
$\delta^{\rm R}(y)$ via
\beq
\label{5.3}
\delta \psi (y)=-\sqrt{1-3y}\, \delta^{\rm R}(y)\,.
\eeq 
Actually,  $\delta \psi^{\rm Dolan}=q\, \delta \psi^{\rm here}$. [Beware that Ref. \cite{Dolan:2013roa} denotes $m_2$ as $M$ and $m_1$ as $\mu$, which conflicts with the normal PN/EOB notation $M=m_1+m_2$, $\mu=m_1 m_2/(m_1+m_2)$ used here.]

Inserting our 8.5PN-accurate result for $\delta^{\rm R}(y)$ in Eq. (\ref{5.3}), and re-expanding in powers of $y$, yields
\begin{widetext}
\begin{eqnarray}
\label{5.4}
\delta\psi^{8.5}(y)&=& y^2-3 y^3-\frac{15}{2} y^4+\left(-\frac{6277}{30}-\frac{496}{15}\ln(2)-8\ln(y)-16\gamma+\frac{20471}{1024}\pi^2\right)y^5\nonumber\\
&&+\left(-\frac{87055}{28}+\frac{653629}{2048}\pi^2-\frac{52}{5}\gamma-\frac{26}{5}\ln(y)+\frac{3772}{105}\ln(2)-\frac{729}{14}\ln(3)\right) y^6-\frac{26536}{1575}\pi y^{13/2}\nonumber\\
&&+\left(-\frac{728644658808461}{91662070500}+\frac{4556}{21}\ln(2)+\frac{3814}{21}\ln(y)+\frac{7628}{21}\gamma+\frac{297761947}{393216}\pi^2+\frac{12879}{35}\ln(3)-\frac{1407987}{524288}\pi^4\right)y^7\nonumber\\
&&-\frac{113411}{22050}\pi y^{15/2}\nonumber\\
&&+\left(\frac{340681718}{1819125}\ln(2)-\frac{74909462}{70875}\gamma+\frac{10374481677311}{22295347200}\pi^2-\frac{199989}{352}\ln(3)-\frac{160934764317}{335544320}\pi^4\right.\nonumber\\
&& +\frac{869696}{1575}\ln(2)\gamma+\frac{1333898219722000637053}{32872539188238750}-\frac{9765625}{28512}\ln(5)-\frac{1344}{5}\zeta(3)+\frac{58208}{105}\ln(2)^2+\frac{3424}{25}\gamma^2\nonumber\\
&& \left.-\frac{37454731}{70875}\ln(y)+\frac{856}{25}\ln(y)^2+\frac{434848}{1575}\ln(y)\ln(2)+\frac{3424}{25}\gamma\ln(y)\right) y^8\nonumber\\
&& +\frac{1179591206}{3274425}\pi y^{17/2}+O_{\ln{}}(y^9)\,.
\end{eqnarray}
\end{widetext}
The first two terms on the right hand side of Eq. (\ref{5.4}), namely the 3PN-accurate expressions $y^2-3y^3$ were previously obtained in Ref. \cite{Dolan:2013roa} from published PN-expanded spin-orbit results \cite{Damour:2007nc,Hartung:2013dza,Marsat:2012fn, Bohe:2012mr}.
[As we shall discuss below, these first two terms can also be easily derived from the EOB formulation of the previously known PN-expanded spin-orbit interaction \cite{Nagar:2011fx,Barausse:2011ys}.] All the  further terms in the PN expansion (\ref{5.4}) are analytically derived here for the first time. Among them, let us note that our analytic results for the coefficient of $y^4$, namely $-15/2$, agrees with the suggestion made in Ref.  \cite{Dolan:2013roa} (from a comparison to their numerical data) that \lq\lq the next (yet unknown) term at 4PN order is close to $-15/2$." 

\begin{figure} 
\typeout{*** EPS figure 1}
\begin{center}
\includegraphics[scale=0.4]{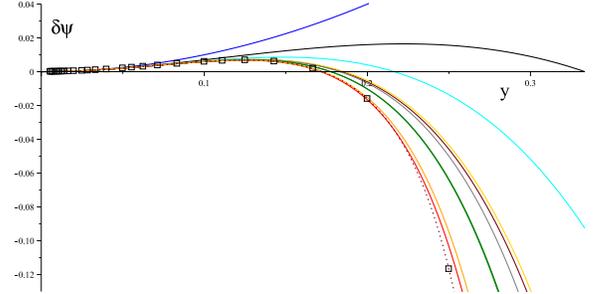}  
\end{center}
\caption{The quantity $\delta \psi$ either from numerical relativity or from PN theory is plotted as a function of $y\in [0,1/3[$. The boxes are the points taken from the numerical results of \cite{Dolan:2013roa} while the dotted curve corresponds to the  fit (\ref{5.11}). The other   curves are the PN approximants of order $2\le n \le 8.5$. The curves are distinguishable from their values when $y\ge 0.2$ where they are ordered from top to bottom as follows: 2PN, 3PN, 4PN, 6PN, 5PN, 6.5PN, 7PN, 7.5PN, 8.5PN, 8PN.}   
\label{fig:1}
\end{figure}

In Fig.~\ref{fig:1}, we compare the successive terms in our analytic PN expansion (\ref{5.4}) of $\delta \psi(y)$ to the numerical results of \cite{Dolan:2013roa}.
It is interesting to note that, while the 3PN-accurate approximation  $\delta \psi^{3\rm PN}(y)=y^2-3y^3$ is unable to reveal the global shape of $\delta \psi(y)$ (with a positive maximum followed by a plunge towards negative values), the higher-order PN approximants are able to qualitatively capture such a strong field behavior. In particular, starting with the 4PN level, the successive PN approximants all suggest a change of sign of $\delta \psi(y)$ in the  strong field regime (around the last stable orbit, $y_{\rm LSO}=1/6$). More precisely, the successive PN approximants to $\delta \psi(y)$ vanish at the  values of $y$ listed in Table I.
\begin{table}[h]
  \caption{\label{tab:pnorder}}
  \begin{center}
    \begin{ruledtabular}
      \begin{tabular}{cc}
PN level & zero of $\delta \psi(y)$\\
\hline
4PN & .2163331\cr 
5PN & .1844183\cr 
6PN & .1858805\cr 
6.5PN &.1830276\cr 
7PN & .1775745\cr 
7.5PN &.1774588\cr 
8PN & .1713716\cr 
8.5PN &.1723855\\
\hline
Numerical (from fit (\ref{5.11})) & .1725217 \cr
 \end{tabular}
  \end{ruledtabular}
\end{center}
\end{table}

Fig. \ref{fig:1} shows that, at least up to $y=0.25$, the successive PN approximants seem to approach better and better the  numerical estimate 
$\delta \psi^{\rm num}(y)$. [See, however, below for the $y>0.25$ domain.] Though the convergence looks monotonic on  Fig. \ref{fig:1}, the values of the zeros listed in table I show that it is actually not monotonic. To gauge the quality of the representation of $\delta \psi^{\rm num}(y)$ by means of our highest PN approximant $\delta \psi^{8.5}(y)$ we compare in Table II the values of $\delta \psi^{\rm num}(y)$ and $\delta \psi^{8.5}(y)$. In particular, the fourth column
lists the base-10 logarithm of the difference $|\delta \psi^{8.5}-\delta \psi^{\rm num}|$. These (logarithmic) differences should be compared with the corresponding numerical errors (as estimated in \cite{Dolan:2013roa}), which are listed in the last column. Note that, for the first twelve values of $y$ (i.e., $y\le \frac{1}{35}$), the difference 
$|\delta \psi^{8.5}-\delta \psi^{\rm num}|$ is smaller than or comparable to the numerical error on $\delta \psi^{\rm num}$. In other words, for these values of $y$, the 8.5PN expression gives, by itself, an excellent fit to the numerical data (with a reduced $\chi^2$ of order of unity). Note also that, up to $y=\frac15$, 
$|\delta \psi^{8.5}-\delta \psi^{\rm num}|$ remains approximatively smaller that $10^{-3}$, which is essentially indistinguishable from zero on a plot such as Fig. \ref{fig:1}. It is only for the last (strong-field) point $y=1/4$ that one can notice a visible discrepancy between $\delta \psi^{8.5}(y)$ and $\delta \psi^{\rm num}(y)$. To check the correctness of the coefficients of our PN expansion, we also evaluated the compatibility between $|\delta \psi^{8.5}-\delta \psi^{\rm num}|$ and the expected, PN remainder $O_{\ln{}}(y^9)$. To do that we have computed the ratio $\Delta (y)=|\delta \psi^{8.5}(y)-\delta \psi^{\rm num}(y)|/(3y)^9$ as a function of $y$.
Here, the conventional inclusion of the factor $3^9$ in the $9$PN level is done to eliminate the expected growth of the $n^{th}$ PN coefficient with $n$ (namely $c_n\sim 3^n$, in view of the pole behavior at the light-ring discussed below). We did find that $\Delta (y)$ remained (for the $y$-values where the numerical error on $\delta \psi^{\rm num}(y)$ is smaller than $\sim (3y)^9$) of order of unity, with only a small variation probably explainable by the expected presence of a logarithmic dependence in the remainder term $O_{\ln{}}(y^9)$.

\begin{widetext}

\begin{table}[t]
  \caption{\label{tab:pnorder}}
  \begin{center}
    \begin{ruledtabular}
      \begin{tabular}{ccccc}
$y$& $ \delta\psi^{8.5\rm PN}$& $\delta\psi^{\rm NR}$   & $\log |\delta\psi^{8.5\rm PN}-\delta\psi_{NR}|$ & $\log |{\rm Err}(\delta\psi_{NR}) |$ \\
\hline
$\frac1{180}$ & .30342{6}4\, $10^{-4}$&  .30342(3)\, $10^{-4}$   &{ $-$9.1959229} &$-$8.5\cr
$\frac1{160}$& .38318{6}0\, $10^{-4}$ &   .38318(1)\, $10^{-4}$  &{ $-$9.2204210} & $-$9.0\cr 
$\frac1{140}$&.499{0}751\, $10^{-4}$ &  .49910(2)\, $10^{-4}$   &$-$8.6035044  & $-$8.7\cr 
$\frac1{120}$ & .676719{4}\, $10^{-4}$ &  .676719(3)\, $10^{-4}$  &{ $-$10.4305158} & $-$9.5\cr 
$\frac1{100}$ & .969242{9}\, $10^{-4}$ & .969242(6)\, $10^{-4}$  &{ $-$10.051832} &  $-$9.2\cr 
$\frac1{90}$ & .119225{9}\, $10^{-3}$ &  .119225(1)\, $10^{-3}$  &{ $-$9.0437309} & $-$9.0\cr
$\frac1{80}$ &.150204{8}\, $10^{-3}$&  .150204(1)\, $10^{-3}$    &{ $-$9.0962689} & $-$9.0\cr 
$\frac1{70}$ & .1950170\, $10^{-3}$ &   .1950169(3)\, $10^{-3}$  &{$-$10.1996486} & $-$9.5\cr 
$\frac1{60}$ &.2632957\, $10^{-3}$ &  .2632957(4)\, $10^{-3}$   &{$-$10.7909356} & $-$9.40\cr 
$\frac1{50}$ &.37475{9}1\, $10^{-3}$ &  .3747588(3)\, $10^{-3}$    &{$-$9.4515689} &$-$9.5\cr 
$\frac1{40}$ & .575052{2}\, $10^{-3}$&  .5750523(3)\, $10^{-3}$   &$-$9.7102148 &  $-$9.5\cr 
$\frac1{35}$ &.741056{2}\, $10^{-3}$ &.7410563(1)\, $10^{-3}$     &$-$10.1569346 & $-$10.0\cr 
$\frac1{30}$ & .99003{1}4\, $10^{-3}$ & .9900329(4)\, $10^{-3}$   &$-$8.8098153 & $-$9.4\cr 
$\frac1{25}$ &.13868{5}6\, $10^{-2}$ &  .13868631(3)\, $10^{-2}$    &$-$8.1500391  & $-$9.5\cr 
$\frac1{20}$ &.2071{4}64\, $10^{-2}$&    .20715008(2)\, $10^{-2}$   &$-$7.4396022 &  $-$9.7\cr 
$\frac1{18}$ & .2488{6}55\, $10^{-2}$ &.24887335(2)\, $10^{-2}$    &$-$7.1032185 & $-$9.7\cr 
$\frac1{16}$ & .3036{5}26\, $10^{-2}$ &  .30367142(1)\, $10^{-2}$   &$-$6.7262699 &  $-$10.0\cr 
$\frac1{14}$ &.376{5}551\, $10^{-2}$ &  .37660516(1)\, $10^{-2}$   &$-$6.3001721 &$-$10.0\cr 
$\frac1{12}$ & .473{3}235\, $10^{-2}$ &  .47347769(2)\, $10^{-2}$   &$-$5.8120760 &$-$9.7\cr
$\frac1{10}$ &.593{2}897\, $10^{-2}$&   .59385649(3)\, $10^{-2}$   &$-$5.2465400 &  $-$9.5\cr 
$\frac19$ &.65{0}8814\, $10^{-2}$ &   .65220522(1)\, $10^{-2}$     &$-$4.8781779 & $-$10.0\cr
$\frac18$ &.6{7}92674\, $10^{-2}$ &    .68178260(4)\, $10^{-2}$     &$-$4.5994355 & $-$9.4\cr
$\frac17$ &.60{3}9472\, $10^{-2}$ & .60923305(9)\, $10^{-2}$        &$-$4.2768883 & $-$9.0\cr
$\frac16$ & .18{1}0206\, $10^{-2}$&   .18780855(8)\, $10^{-2}$      & $-$4.1682628 &  $-$9.1\cr
$\frac15$ &-.1{5}37240\, $10^{-1}$&    $-$.16055004(5)\, $10^{-1}$     &  $-$3.1658286 & $-$8.3\cr
$\frac14$ &$-$.{9}478438\, $10^{-1}$& $-$.11669057(3)       & $-$1.6594332 & $-$7.5\cr 
   \end{tabular}
  \end{ruledtabular}
\end{center}
\end{table}
\end{widetext}

A striking feature of the shape of $\delta \psi^{\rm num}(y)$ is its plunging behavior, towards negative values, for $y\ge 0.2$. Let us now show how this behavior can be understood essentially analytically. Let us first recall that Akcay et al. \cite{Akcay:2012ea} have shown both analytically and numerically, that (barring special cancellations) all the components (in Lorenz gauge) of the regularized metric $h_{\mu\nu}^{\rm R}(x^\lambda)$ will blow up proportionally to $(1-3y)^{-1/2}$ as one approaches the light-ring, $r_0=3m_2$, corresponding to $y=m_2/r_0+O(q)=\frac13 +O(q)$. [This follows simply from the fact that the distributional source $T^{\mu\nu}$ associated with the moving mass $m_1$ contains a singular factor $(1-3y)^{-1/2}$ when $y\to 1/3$.]
Among the various terms of the explicit expression (\ref{3.8}) of $\delta(y)$ we notice that there is the term
\beq
\label{5.5}
-\frac12 \frac{y}{(1-2y)(1-3y)}h_{kk}^{\rm R}\,,
\eeq
which is the only one to contain a prefactor that is singular on the light-ring. The other terms contain the product of regular prefactors with a component $h_{\mu\nu}^{\rm R}$
or $\partial_\lambda h_{\mu\nu}^{\rm R}$.
By the reasoning recalled above, the latter terms are expected to blow-up at most $\propto (1-3y)^{-1/2}$ as $r_0$ approaches the light-ring. This leaves the single term (\ref{5.5}) as the most singular contribution near $y=1/3$, of order $(1-3y)^{-3/2}$. Ref. \cite{Akcay:2012ea} studied the singular behavior of the quantity $h_{uu}^{R,L}=\Gamma^2 h_{kk}^{R,L}$ (where the extra superscript L refers to the Lorenz gauge), and found that it behaved as
\beq
\label{5.6}
h_{uu}^{R,L}(y)\approx -\frac12 \zeta (1-3y)^{-3/2}\,,
\eeq
with a numerical factor $\zeta \approx 1$. [More precisely, from Table II in \cite{Akcay:2012ea} we see $\zeta=1.006(3)$.] Using $\Gamma^2=1/k^2=(1-3y)^{-1}+O(q)$ and the link given by Eq. (17) in \cite{Akcay:2012ea} between the Lorenz-gauge value of $h_{uu}^{\rm R}$ and its (gauge-invariant) value in an asymptotically flat gauge, one easily derives that, in  such a \lq\lq flat" gauge (as used here)
\beq
\label{5.7}
h_{kk}^{\rm R}(y)\approx \left(-\frac{\zeta}{2}+\frac29 \right) (1-3y)^{-1/2}\,.
\eeq
Inserting this result in Eq. (\ref{5.5}) predicts that the leading-order singularity of $\delta(y)$ as $y\to 1/3$ will be
\beq
\label{5.8}
\delta(y)\approx +\frac14\left(\zeta-\frac49 \right) (1-3y)^{-1/2}\,.
\eeq
In turn, this behavior predicts that $\delta \psi(y)$ will blow-up near the light-ring as
\beq
\label{5.9}
\delta \psi(y) \approx -\frac14\left(\zeta-\frac49 \right) (1-3y)^{-1}\,.
\eeq
In other words, the results of \cite{Akcay:2012ea}, together with our explicit expression (\ref{3.8}), predict the behavior  (\ref{5.9}) (i.e., a pole-like plunge towards $-\infty$), with a numerical coefficient
\beq
\label{5.10}
 -\frac14\left(\zeta-\frac49 \right)\approx -0.1404(1)\,.
\eeq
This prediction is clearly qualitatively compatible with the numerical results of \cite{Dolan:2013roa}. It suggests ways of defining global fits for the strong field behavior $\delta \psi (y)$ by a priori incorporating the pole-like behavior (\ref{5.9}). We have indeed found that one gets a rather good fit to the numerical data $\psi^{\rm num}(y)$ by fitting it to the following simple form
\beq
\label{5.11}
\psi^{\rm fit}(y)= \frac{y^2}{1-3y}\frac{(1-a_0 y)(1+a_1 y +a_2 y^2)}{1+b_1 y}\,.
\eeq
The prefactor $y^2(1-3y)^{-1}$ in Eq. (\ref{5.11}) incorporates the lowest PN information, as well as the existence of a pole
 at $y=1/3$. The factor $(1-a_0y)$ incorporates the existence of a zero in $\psi(y)$. It is also natural to incorporate the 3PN information that
$\psi^{3\rm PN}(y)=y^2-3y^3+O(y^4)$. This yields the constraint
\beq
\label{5.12}
b_1=6 +a_1-a_0\,.
\eeq
Using the latter constraint, we have only three arbitrary parameters to be fitted. By doing a simple (unweighted) least-square fit we found a rather good fit to all existing data with  
\begin{eqnarray}
\label{5.13}
a_0&=& 5.7963711\,,\nonumber\\
a_1&=& -2.1239228\,,\nonumber\\
a_2&=& 1.1418178\,,
\end{eqnarray}
and with $b_1$ computed from Eq. (\ref{5.12}).
The unweighted residual standard deviation of this fit is $0.51 \times  10^{-6}$, while its maximum absolute deviation from the numerical data is $1.21\times  10^{-6}$.
Such a fit is probably sufficiently accurate for most practical uses. It is plotted in Fig. \ref{fig:1} both to interpolate between numerical data points and to extrapolate them in the very strong-field domain $0.25\le u \le 1$.

Let us comment on the information that can be extracted from our fit (\ref{5.11})-(\ref{5.13}). First, the fitted value of $a_0$ yields an accurate estimate of the radius $r_0$ at which $\delta \psi$ vanishes, namely $r_0=5.7963711m_2$. [This is close to the value $r_0\approx 5.8 m_2$ estimated in  \cite{Dolan:2013roa}]. Second, the fit (\ref{5.11}) implies that the coefficient of $(1-3y)^{-1}$ in the asymptotic behavior (\ref{5.9}) of $\delta \psi (y)$ near the light-ring is numerically equal to
\beq
\label{5.14}
\lim_{y\to 1/3}(1-3y)\delta \psi^{\rm fit}(y)\approx -.1205456\,.  
\eeq
Note that this is rather close to the value (\ref{5.10})  predicted from the results of Ref. \cite{Akcay:2012ea}.
Evidently, as the last numerical data point is located at $y=1/4$, we cannot expect to accurately reproduce the value (\ref{5.10}) which used a value of $\zeta$ extracted from data on $h_{kk}$ much closer to the light-ring. It would be interesting to compute $\delta \psi(y)$ for values of $y$ close to $1/3$, so as to investigate with more accuracy the link (\ref{5.9}).

\section{Gyro-gravitomagnetic ratios and effective-one-body reformulation}

At linear order in both spins, the Hamiltonian of a system of two spinning bodies reads (with $a,b=1,2$ being body labels; and suppressing the superscript \lq\lq can" on the spin vectors)
\beq
\label{6.1}
H({\mathbf x}_a,{\mathbf p}_a,{\mathbf S}_a )=H_{\rm (orb)}({\mathbf x}_a,{\mathbf p}_a)+H_{\rm SO}({\mathbf x}_a,{\mathbf p}_a,{\mathbf S}_a )\,,
\eeq
with
\beq
\label{6.2}
H_{\rm SO}({\mathbf x}_a,{\mathbf p}_a,{\mathbf S}_a )=\sum_{a=1,2}\bfOmega_a ({\mathbf x}_b,{\mathbf p}_b)\cdot {\mathbf S}_a\,.
\eeq
We recall that the spin-orbit coupling coefficient $\bfOmega_a ({\mathbf x}_b,{\mathbf p}_b)$ (which coincides with the vectorial precession frequency of ${\mathbf S}_a$
with respect to the time $t$ used in the Hamiltonian formulation) have been determined, within PN theory, at increasing PN accuracies (up to the 3PN level) in Refs \cite{Faye:2006gx,Blanchet:2006gy,Damour:2007nc,Steinhoff:2007mb,Steinhoff:2008zr,Steinhoff:2008ji,Porto:2008tb,Porto:2008jj,Porto:2010tr,Levi:2010zu,Hartung:2011te,Hartung:2013dza,Marsat:2012fn,Bohe:2012mr}. 
When using PN theory, one can determine the exact dependence of $\bfOmega_a$ on the two masses of the system, $m_1$ and $m_2$. Note that in this respect it is enough to  know $\bfOmega_1$ as a function of  ${\mathbf x}_a$, ${\mathbf p}_a$ and $m_a$ because $\bfOmega_2$ is then obtained by exchanging the two particle labels $1$ and $2$:
\beq
\label{6.3}
\bfOmega_2 ({\mathbf x}_1,{\mathbf x}_2,{\mathbf p}_1,{\mathbf p}_2,m_1,m_2)=
\bfOmega_1 ({\mathbf x}_2,{\mathbf x}_1,{\mathbf p}_2,{\mathbf p}_1,m_2,m_1)
\,.
\eeq
When working in the center-of-mass frame (${\mathbf p}_1+{\mathbf p}_2={\mathbf 0}$) one can express the Hamiltonian in terms of the relative-motion canonical variables
${\mathbf R}={\mathbf x}_1-{\mathbf x}_2$, ${\mathbf P}={\mathbf p}_1=-{\mathbf p}_2$.
One then finds that
\beq
\label{6.4}
H_{\rm SO}=\frac{G}{c^2R^3} {\mathbf L}\cdot (g_{S_1}{\mathbf S}_1+g_{S_2}{\mathbf S}_2)
\,,
\eeq
where ${\mathbf L}={\mathbf R}\times {\mathbf P}$, and where
$g_{S_1}$ is a dimensionless quantity which depends on $GM/(c^2R)$, ${\mathbf P}^2/\mu^2$, $P_R^2/\mu^2$, $m_1$ and $m_2$ (with the symmetry property $g_{S_2}(m_1,m_2)=g_{S_1}(m_2,m_1)$.  
At lowest order in PN expansion the values of these dimensionless coefficients are
\beq
g_{S_1}=2 +\frac{3}{2}\frac{m_2}{m_1}+O\left(\frac{1}{c^2}\right), \qquad
g_{S_2}=2 +\frac{3}{2}\frac{m_1}{m_2}+O\left(\frac{1}{c^2}\right)\,.
\eeq
For many purposes, as the full Hamiltonian (\ref{6.1}) is symmetric under the particle label exchange $1\leftrightarrow 2$, it is convenient to replace the individual \lq\lq gyro-gravitomagnetic ratios" $g_{S_1}$ and $g_{S_2}$ by two new $1$-$2$-symmetric gyro-gravitomagnetic ratios  $g_{S}$ and $g_{S_\ast}$
defined (following \cite{Damour:2008qf}) such that
\beq
\label{6.5}
H_{\rm SO}=\frac{G}{c^2R^3} {\mathbf L}\cdot (g_{S}{\mathbf S}+g_{S_\ast}{\mathbf S}_\ast)\,,
\eeq
where ${\mathbf S}$ and ${\mathbf S}_\ast$ are the following two basic symmetric combinations of the spin vectors
\begin{eqnarray}
\label{6.6}
{\mathbf S}&=& {\mathbf S}_1+{\mathbf S}_2\,,\ \qquad
{\mathbf S}_\ast =\frac{m_2}{m_1}{\mathbf S}_1+\frac{m_1}{m_2}{\mathbf S}_2\,,
\end{eqnarray}
with inverse relations
\begin{eqnarray}
\label{6.6bis}
{\mathbf S}_1&=&\frac{m_1}{m_1^2-m_2^2}(m_1 {\mathbf S}-m_2 {\mathbf S}_\ast)\,,\nonumber\\
{\mathbf S}_2&=&-\frac{m_2}{m_1^2-m_2^2}(m_2 {\mathbf S}-m_1 {\mathbf S}_\ast)\,.
\end{eqnarray}
Therefore, 
\begin{eqnarray}
\label{6.7}
g_{S_1}&=& g_S+\frac{m_2}{m_1}g_{S_\ast}\nonumber\\
g_{S_2}&=& g_S+\frac{m_1}{m_2}g_{S_\ast}\,,
\end{eqnarray}
or, equivalently,
\begin{eqnarray}
\label{6.8}
g_{S}&=& \frac{m_1^2}{m_1^2-m_2^2}g_{S_1}-\frac{m_2^2}{m_1^2-m_2^2}g_{S_2}\nonumber\\
g_{S_\ast}&=& -\frac{m_1m_2}{m_1^2-m_2^2}(g_{S_1}-g_{S_2})\,.
\end{eqnarray}
Note that to lowest PN order the values of these \lq\lq symmetric" gyro-gravitomagnetic ratios are
\beq
g_{S}= 2+O\left(\frac{1}{c^2} \right)\,,\qquad 
g_{S_\ast}=\frac32+O\left(\frac{1}{c^2} \right)\,.
\eeq

A serious inconvenient of the gyro-gravitomagnetic ratios $g_S$ and $g_{S_\ast}$ is that they are gauge-dependent: they depend on the choice of phase-space coordinates, and notably on the choice of the radial coordinate $R$ appearing
(cubed)  as an overall factor in Eq. (\ref{6.5}). This gauge dependence is alleviated when reformulating the Hamiltonian $H$ within the EOB formalism. In that formalism, the Hamiltonian is first rewritten in terms of an auxiliary \lq\lq effective" Hamiltonian $H_{\rm eff}$, such that
\beq
\label{6.9}
H\equiv Mc^2 \sqrt{1+2\nu \left(\frac{H_{\rm eff}}{\mu c^2} -1 \right)}\,.
\eeq
This is equivalent to
\begin{eqnarray}
\label{6.10}
\frac{H_{\rm eff}}{\mu c^2} &=& \frac{H^2-m_1^2c^4-m_2^2c^4 }{2m_1m_2c^4}\nonumber\\
&=&1+\frac{H^{\rm nr}}{\mu c^2}+\frac12 \nu \left(\frac{H^{\rm nr}}{\mu c^2}\right)^2\,.
\end{eqnarray}
where $H^{\rm nr}=H-(m_1+m_2)c^2$ denotes the \lq\lq non-relativistic" part  of the total Hamiltonian.

Note that Eq. (\ref{6.10}) is exact (both in a PN sense and in a GSF sense). At linear order in the spins, the replacement $H=H_{\rm orb}+H_{\rm SO}$ in Eq. (\ref{6.10}) yields, when allowing also for a spin-dependent canonical transformation (with generating function $G_s$) between the original phase-space coordinates and the EOB ones, 
\begin{eqnarray}
\label{6.11}
&&H^{\rm eff}({\mathbf R}^{\rm EOB}, {\mathbf P}^{\rm EOB}, {\mathbf S}_1, {\mathbf S}_2)=H_{\rm orb}({\mathbf R}^{\rm EOB}, {\mathbf P}^{\rm EOB})\nonumber\\
&&\qquad\quad
+H^{\rm eff}_{\rm SO}({\mathbf R}^{\rm EOB}, {\mathbf P}^{\rm EOB}, {\mathbf S}_1, {\mathbf S}_2)\,,
\end{eqnarray}
with
\beq
\label{6.12}
H^{\rm eff}_{\rm SO}=\frac{H^{\rm eff}_{\rm orb}}{Mc^2}\left(H_{\rm SO} +\{G_s,H_{\rm orb}\}  \right)\,,
\eeq
so that the effective \lq\lq gyro-gravitomagnetic" ratios $g_S^{\rm eff}$ and $g_{S^\ast}^{\rm eff}$ entering the effective Hamiltonian, i.e.,
\beq
\label{6.13}
H^{\rm eff}_{\rm SO}= \frac{G}{c^2 R_{\rm EOB}^3}{\mathbf L} \cdot (g_S^{\rm eff}{\mathbf S}+g_{S^\ast}^{\rm eff}{\mathbf S}^\ast)\,,
\eeq
differ from the ratios $g_S$ and $g_{S^\ast}$ entering Eq. (\ref{6.5}) by a common factor $(H_{\rm orb}/(Mc^2))(R_{\rm EOB}/R)^3$, and by an extra term involving the Poisson bracket $\{G_s, H_{\rm orb}  \}$. The effective gyro-gravitomagnetic ratios have PN expansions of the form
\begin{eqnarray}
\label{6.14}
g_S^{\rm eff}&=& 2+\frac1{c^2}g_{S}^{(2)}\left({\mathbf P}^2,P_R^2,\frac{GM}{R}; a_0\right)\nonumber\\
&+& \frac1{c^4}g_{S}^{(4)}\left({\mathbf P}^2,P_R^2,\frac{GM}{R}; a_0,a_1,a_2,a_3\right)+\ldots \nonumber\\
g_{S^\ast}^{\rm eff}&=&  \frac32 +\frac1{c^2}g_{S^\ast}^{(2)}\left({\mathbf P}^2,P_R^2,\frac{GM}{R}; b_0\right)\nonumber\\
&+&\frac1{c^4}g_{S^\ast}^{(4)}\left({\mathbf P}^2,P_R^2,\frac{GM}{R}; b_0,b_1,b_2,b_3\right)+\ldots
\end{eqnarray}
where the $a_n$'s and $b_n$'s are spin-gauge parameters related to a remaining arbitrariness in the rotational state of the frame with respect to which the spin components are measured \cite{Damour:2008qf,Nagar:2011fx,Barausse:2011ys}. However, as pointed out in Ref. \cite{Damour:2008qf}, in the case of circular orbits all the gauge freedom in $g_S^{\rm eff}$ and $g_{S^\ast}^{\rm eff}$ disappear, and the two \lq\lq circular" gyro-gravitomagnetic ratios $g_S^{\rm eff,circ}$ and $g_{S^\ast}^{\rm eff, circ}$ become gauge-independent functions of the gauge-independent variable $u=GM/(c^2R_{EOB})$. [Indeed, $u$ is uniquely defined by the gauge-fixing used in defining the orbital part of the EOB Hamiltonian. See, e.g below for the unambiguous link between $u$ and other gauge-invariant quantities such as $|{\mathbf L}|$ and $\Omega$.]

The current (PN and GSF) knowledge of these functions is
\begin{eqnarray}
\label{6.15}
g_S^{\rm eff,circ}&=&g_{S (0)}^{\rm eff,circ}(u)-\frac{5}{8}\nu u -\left(\frac{51}{4}\nu +\frac18 \nu^2 \right)u^2\nonumber\\
&&  +\nu O(u^3) \nonumber\\
\label{6.16}
g_{S^\ast}^{\rm eff, circ}&=&  g_{S^\ast (0)}^{\rm eff,circ}(u)-\frac{3}{4}\nu u -\left(\frac{39}{4}\nu +\frac{3}{16} \nu^2 \right)u^2 \nonumber\\
&& +\nu O(u^3)\,,
\end{eqnarray}
where
\beq
\label{6.17}
g_{S(0)}^{\rm eff,circ}(u)=2
\eeq
and \cite{Barausse:2009aa}
\beq
\label{6.18}
 g_{S^\ast (0)}^{\rm eff,circ}(u)=\frac{3}{1+\frac{1}{\sqrt{1-3u}}}=\frac32 -\frac{9}{8}u -\frac{27}{16}u^2-\ldots 
\eeq
Here the quantities $g_{S(0)}^{\rm eff,circ}$ and $ g_{S^\ast(0)}^{\rm eff,circ}$ denote the exact test-mass limits $(\nu \to 0)$ of the gyro-gravitomagnetic ratios. Note that while $g_S{}_{(0)}^{\rm eff,circ}(u)$ is a constant, $ g_{S^\ast}{}_{(0)}^{\rm eff,circ}(u)$ is a function of $u$ which decreases from the infinite-separation value $\lim_{u\to 0}g_{S^\ast}{}_{(0)}^{\rm eff,circ}(u)=3/2$ to $g_{S^\ast}{}_{(0)}^{\rm eff,circ}(1/3)$=0 at the light-ring.

Using the above results, one can easily translate our results on $\delta(y)$ into an improved knowledge of $g_{S^\ast}^{\rm eff, circ}$.
First, we note that when ${\mathbf S}_2={\mathbf 0}$, and when  ${\mathbf S}_1$ is simply taken to be parallel to the orbital angular momentum ${\mathbf L}={\mathbf R}\times {\mathbf P}$, the spin-orbit interaction energy reads (taking into account the vanishing of $\{G_s,H_{\rm orb}\}$ along circular orbits)
\begin{eqnarray}
\label{6.19}
H_{SO}&=&\Omega_1S_1\\
&=&G \frac{M}{H_{\rm orb}}\frac{L}{R_{\rm EOB}^3}\left(g_S^{\rm eff,circ}(u)+\frac{m_2}{m_1}g_{S^\ast}^{\rm eff,circ}(u) \right)S_1\,.\nonumber
\end{eqnarray}
Working with the dimensionless EOB variables $u=GM/(c^2R_{\rm EOB})$, $l=L/(GM\mu)$, this yields
\begin{eqnarray}
\label{6.20}
&&(m_1+m_2)\Omega_1=\frac{l(u) u^3}{\sqrt{1+2\nu (\hat H_{\rm eff}(u)-1)}}\times \nonumber\\
&& \quad\left(\nu g_S^{\rm eff,circ}(u)+\frac{m_2}{M}g_{S^\ast}^{\rm eff,circ}(u)\right)\,,
\end{eqnarray}
where the EOB functions of $u$ appearing on the right hand side are explicitly given by
\begin{eqnarray}
\label{6.21}
l(u)&=& \sqrt{-\frac{A'(u)}{(u^2A(u))'}}\nonumber\\
\label{6.22}
\hat H_{\rm eff}(u) &=& \sqrt{A(u)(1+l^2(u) u^2)}\nonumber\\
&=&\frac{A(u)}{\sqrt{A(u)+\frac12 u A'(u)}}
\end{eqnarray}
Note that the the right hand side of Eq. (\ref{6.20}) is expressed as a function of $u$, while the left hand side is known as a function of $y=(m_2\Omega)^{2/3}$, namely
\begin{widetext}
\begin{eqnarray}
\label{6.23}
(m_1+m_2)\Omega_1(y)=\left(1+\frac{m_1}{m_2}  \right)y^{3/2}
\left[
  1-\sqrt{1-3y}\left( 
1+\frac{m_1}{m_2}\delta^{\rm R}(y)+O\left( \frac{m_1}{m_2} \right)^2 
\right)  
\right]\,.
\end{eqnarray}
\end{widetext}
The link between $u$ and $y$, or better, between $u$ and 
\beq
\label{6.24}
x=((m_1+m_2)\Omega)^{2/3}=\left(1+\frac{m_1}{m_2}\right)^{2/3}y
\eeq
is provided by the EOB result \cite{Damour:2009sm}
\beq
\label{6.25}
x(u)=u \left( \frac{-\frac12 A'(u)}{1+2\nu (\hat H_{\rm eff}^{\rm circ}(u)-1)} \right)^{1/3}\,.
\eeq
If we re-express $m_2 \Omega_1$ as a function of $x$, we find, at linear order in $q$,
\begin{eqnarray}
\label{new_6.27}
&&m_2\Omega_1(x)= x^{3/2}(1-\sqrt{1-3x})-qx^{3/2}\times \nonumber\\
&&\quad \left[1-\sqrt{1-3x}(1-\delta^{\rm R}(x))+\frac{x}{\sqrt{1-3x}}  \right]\,.
\end{eqnarray}
Inserting the GSF expansion of the basic EOB radial potential $A(u; \nu)$, i.e.,
\beq
\label{6.26}
A(u)=1-2u+\nu a_{1\rm SF}(u)+O(\nu^2)
\eeq
and expanding all the above results in powers of $q=m_1/m_2$ and/or of $\nu=q/(1+q)^2$, we easily see that the zeroth order term in the GSF expansion of Eq. (\ref{6.20}) (i.e., the limit $q\to 0$ on both sides) yields
\beq
\label{6.27}
 g_{S^\ast(0)}^{\rm eff,circ}(u)=\sqrt{1-3u}\frac{1-\sqrt{1-3u}}{u}=\frac{3\sqrt{1-3u}}{1+ \sqrt{1-3u} }
\eeq 
which is equivalent to Eq. (\ref{6.18}). Note that, at this zeroth order in $q$, the value of the S-type gyro-gravitomagnetic ratio, $g_S^{\rm eff}$, did not matter.

At the first GSF order, i.e., when keeping terms linear in $\nu$ or $q$, one sees from Eq. (\ref{6.15}) that the knowledge $g_S^{\rm eff}=2+O(\nu)$ will be enough to relate the 1SF contribution to $g_{S^\ast}^{\rm eff}$,
\beq
\label{6.28}
 g_{S^\ast}^{\rm eff,circ}(u)= g_{S^\ast(0)}^{\rm eff,circ}(u)+\nu  g_{S^\ast 1\rm SF}^{\rm eff,circ}(u)+O(\nu^2)
\eeq
to $\delta(y)$, or, equivalently to $\delta \psi(y)=-\sqrt{1-3y}\, \delta(y)$.
We find
\begin{eqnarray}
\label{our_res}
g_{S_\ast 1SF}(u) &=& 
\frac{1}{\Gamma_0}\frac{\delta\psi(u)}{u}  +\frac32 \frac{\Gamma_0^2}{\Gamma_0+1}a_{1\rm SF}(u)\nonumber\\
&& +\frac{1}4(\Gamma_0-2)  a_{1\rm SF}'(u)\nonumber\\
&& -\frac13 \frac{(\Gamma_0-1)}{\Gamma_0(\Gamma_0+1)}(\Gamma_0^2+8\Gamma_0-2)\,,
\end{eqnarray}
where we used the shorthand notation $\Gamma_0(u)\equiv (1-3u)^{-1/2}$.
Note the presence of a factor $1/u$ in the coefficient of $\delta \psi (u)$ which  decreases the PN accuracy with which we can compute $g_{S_*}^{1SF}(u)$. [Actually, $\delta \psi(u)=u^2 -3u^3+O(u^4)$, so that $\delta \psi(u)/u=u -3u^2+O(u^3)$.] Similarly, the presence of a derivative acting on $a_{1\rm SF}(u)$ also decreases the  PN accuracy with which we can compute $g_{S_*}^{1SF}(u)$. The last term in Eq. (\ref{our_res}) can be rewritten in terms of $k_0(u)\equiv \sqrt{1-3u}$ as
\begin{eqnarray}
\label{last_term}
&&-\frac13 \frac{(\Gamma_0-1)}{\Gamma_0(\Gamma_0+1)}(\Gamma_0^2+8\Gamma_0-2)=\nonumber\\
&& -\frac23 k_0 + 4 -\frac{1}{3k_0}-\frac{6}{k_0 +1}\,.
\end{eqnarray}

By inserting in Eq. (\ref{our_res}) the PN  expansion, Eq. (\ref{5.4}), of $\delta\psi(u)$ up to $u^{8.5}$, together with the PN expansion of $a_{1\rm SF}(u)$ which was determined in Ref.  \cite{Bini:2014nfa} up to $u^{8.5}$, we obtain [after PN re-expanding $\Gamma_0(u)$] the following PN expansion of $g_{S_*}^{1SF}(u)$ up to $u^{7.5}$
\begin{eqnarray}
g_{S_*}^{1SF}(u)&=&-\frac{3}{4}u-\frac{39}{4}u^2+\left(\frac{41}{32}\pi^2-\frac{7627}{192}\right)u^3\nonumber\\
&& +(g_4^c+g_4^{\ln{}}\ln u) u^4 \nonumber\\
&& +(g_5^c+g_5^{\ln{}}\ln u) u^5 -\frac{93304}{1575}\pi u^{11/2}\nonumber\\
&& +(g_6^c+g_6^{\ln{}}\ln u) u^6 +\frac{4596019}{12600}\pi u^{13/2}\nonumber\\
&& +(g_7^c+g_7^{\ln{}}\ln u+g_7^{\ln^2{}}\ln^2 u) u^7\nonumber\\
&&  +\frac{118299749}{2182950}\pi u^{15/2}\,,
\end{eqnarray}
where
\begin{eqnarray}
g_4^c&=& -\frac{1017}{20}-\frac{1456}{15}\ln 2-48\gamma+\frac{23663}{2048}\pi^2\nonumber\\
g_4^{\ln{}}&=&-24 
\end{eqnarray}
\begin{eqnarray}
g_5^c&=& -\frac{161160813}{89600}+\frac{70696}{105}\ln 2+\frac{9832}{35}\gamma+\frac{712905}{4096}\pi^2\nonumber\\
&&-\frac{729}{7}\ln 3\nonumber\\
g_5^{\ln{}}&=&\frac{4916}{35}
\end{eqnarray}
\begin{eqnarray}
g_6^c&=& -\frac{29750077105462223}{11732745024000}-\frac{674904611}{7077888}\pi^2+\frac{480829}{2835}\gamma\nonumber\\
&& -\frac{2954531}{2835}\ln2+\frac{315657}{280}\ln  3+\frac{16790137}{1048576}\pi^4\nonumber\\
g_6^{\ln{}}&=&\frac{480829}{5670} 
\end{eqnarray}
\begin{eqnarray}
g_7^c&=& \frac{1167584}{525}\ln(2)^2+\left(-\frac{5587843424}{779625}+\frac{499904}{225}\gamma\right)\ln 2 \nonumber\\
&&-\frac{12227517}{3080}\ln 3-\frac{1953125}{3564}\ln 5\nonumber\\
&& +\frac{4143031385722624236137377}{53858368206010368000}\nonumber\\
&& -\frac{204902966117}{335544320}\pi^4-1088\zeta(3)-\frac{1830427308991}{2229534720}\pi^2\nonumber\\
&&+\frac{58208}{105}\gamma^2-\frac{903605468}{121275}\gamma\nonumber\\
g_7^{\ln{}}&=& -\frac{451802734}{121275}+\frac{58208}{105}\gamma \nonumber\\
g_7^{\ln^2{}}&=& \frac{14552}{105}\,.
\end{eqnarray}
For convenience, the (approximated) numerical expression of $g_{S_\ast}^{1\rm SF}$ is also given below
\begin{eqnarray}
g_{S_*}^{1SF}(u)&\approx &-.75 u-9.75 u^2-27.07852769 u^3\nonumber\\
&& +(-24\ln u  -31.8024627) u^4\nonumber\\
&& +(140.4571429\ln u +433.5539014) u^5\nonumber\\
&& -186.1099435 u^{11/2}\nonumber\\
&& +(84.80229277\ln u-1302.962118) u^6\nonumber\\
&& +1145.938058 u^{13/2}\nonumber\\
&& +(-4339.91599-2635.437946\ln u \nonumber\\
&& +138.5904762\ln^2 u ) u^7\nonumber\\
&& +170.2510925 u^{15/2}\,.
\end{eqnarray}

The first two terms of the latter expansion, namely $g_{S_*}^{1SF}(u)= -\frac34 u -\frac{39}{4}u^2+\ldots$, agree with the only previously known terms, which were exhibited in Eq. (\ref{6.15}).
We note that the first three coefficients in the PN expansion of $g_{S_*}^{1SF}(u)$ are negative, thereby continuing the trend noticed in Ref. \cite{Damour:2007nc,Nagar:2011fx} that
the $\nu$-dependent contributions to $g_{S_*}(u; \nu)$ further reduce the value of $g_{S^\ast (0)}(u)$ which was itself decreasing as $u$ was increasing.

It is, however,  interesting to note that this trend is changed because of the behavior of the terms depending on $a_{1\rm SF}'(u)$ in Eq. (\ref{our_res}).
First,  the contribution to  $g_{S_*}^{1SF}(u)$ coming from $\delta \psi(u) $ is initially positive and then becomes negative before plunging towards $-\infty$  like $\approx -\frac{c_\psi}{\sqrt{1-3u}}$ where $c_\psi\approx 0.36$ when $u\to1/3$.
 By contrast, one sees that, as $u\to1/3$, the dominant contribution from the other terms is the one linked to $a_{1\rm SF}'(u)$ which behaves asymptotically as $+\frac{c_{a'}}{(1-3u)^2}$, with a positive   constant $c_{a'}=3\zeta/32 \approx .094$. [The behavior of $a_{1\rm SF}(u)\approx \frac{\zeta}{4\sqrt{1-3u}}$ (where $\zeta \approx 1$) as the light-ring is approached was obtained in \cite{Akcay:2012ea}.]
This asymptotic behavior near the light-ring shows that $g_{S_*}^{1SF}(u)$ will change sign near the light-ring to become positive.

\begin{figure} 
\typeout{*** EPS figure 2}
\begin{center}
\includegraphics[scale=0.3]{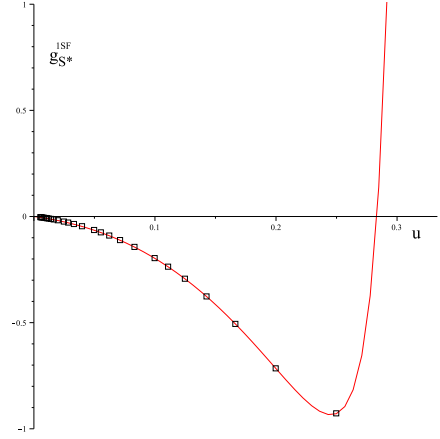}     
\end{center}
\caption{The quantity $g_{S_\ast}^{1 \rm SF}$ either exact from numerical relativity or approximated by using the data fits is plotted as a function of  $u\in [0,1/3]$.
The used data fits are obtained from our model (\ref{5.11}) for $\delta \psi$, and from model $\# 14$ of Ref. \cite{Akcay:2012ea} for $a_{1\rm SF}(y)$.}  
\label{fig:gS1star}
\end{figure}

By using our fitting function (\ref{5.11})  for $\delta \psi(u)$ and the fitting function of Ref. \cite{Akcay:2012ea} for $a_{1\rm SF}(u)$ we have analyzed the behavior of $g_{S_\ast }^{1SF}(u)$, determining the location of the zero at $u\approx .2833343$.
By using instead the numerical relativity data points for $\delta \psi(u)$ and the fitting function of Ref. \cite{Akcay:2012ea} for $a_{1\rm SF}(u)$ we have extracted
a sequence of \lq\lq numerical" data points for $g_{S_*}^{1SF}(u)$.  Finally, with such informations, together with the beginning of the PN expansion of 
$g_{S_*}^{1SF}(u)$, we have determined the following fitting curve 
\beq
\label{gsstar_fit}
g_{S_\ast }^{1SF}{}^{\rm fit}(u)=- \frac{3u(1-A_0 u) (1+A_1 u+A_2 u^2+A_3 u^3)}{4(1-3u)^2(1+B_1 u^2+B_2 u^3) }\,. 
\eeq
with $A_0=3.5293991=1/ .2833343$ known from the position of the zero of $g_{S_*}^{1SF}(u)$ and $A_1=7+A_0= 10.5293991$ also known from the request that the series expansion of $g_{S_*}^{1SF}{}^{\rm fit}(u)$ should start as $-\frac34 u -\frac{39}{4}u^2+\ldots$. The remaining parameters, obtained by a standard fitting procedure, are listed below:
\begin{eqnarray}
A_2&=& 2.0797445\,,\qquad  A_3= -80.9910909\,,\nonumber\\
B_1&=& 0.4689439\,,\qquad 
B_2=-5.1432878\,.
\end{eqnarray}
They are such that the maximum absolute deviation from the numerical data is $5.25\, 10^{-5}$.
The results are also shown in Fig. \ref{fig:gS1star}. 

\begin{figure} [h]
\typeout{*** EPS figure 3}
\begin{center}
\includegraphics[scale=0.3]{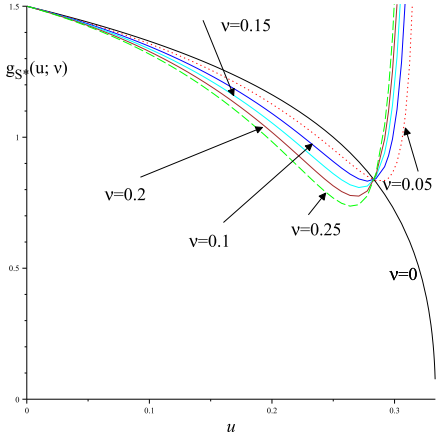}      
\end{center}
\caption{The  full function $g_{S_\ast}^{\rm eff,circ}(u; \nu)$  is plotted as a function of $u$, for selected values of $\nu=[0,0.05,0.1,0.15,0.2,0.25]$. }
\label{fig:3}
\end{figure}

In Fig. \ref{fig:3} we exhibit the $\nu$-dependence of the full function $g_{S_\ast}^{\rm eff,circ}(u; \nu)$ approximated as being the sum of the test-mass limit $g_{S_\ast (0)}(u)$, Eq. (\ref{6.18}), the 1SF-contribution $g_{S_*}^{1SF}{}^{\rm fit}(u)$ computed by means of the fit  (\ref{gsstar_fit}), and the only currently known contribution of order $\nu^2$, namely the last term in Eq. (\ref{6.16}): $-\frac{3}{16}\nu^2u^2$. 
This function is plotted between $u=0$ and $u=\frac{1}{3}$ for the values $\nu=[0,0.05,0.1,0.15,0.2,0.25]$. Note that because of the change of sign of $g_{S_*}^{1SF}(u)$ at $u\approx 0.283$ the various curves for $g_{S_\ast}^{\rm eff,circ}(u; \nu)$ approximately cross the test-mass limit $g_{S_\ast (0)}^{\rm eff,circ}(u)$ around $u\approx 0.283$ before increasing in a divergent manner near the light-ring. As discussed in Refs. \cite{Akcay:2012ea,Damour:2012ky}, the contribution of higher powers of $\nu$ may significantly affect the exact behavior near the light-ring. We leave to future work a discussion of this issue.

Finally, let us also quote the expression of $ g_{S^\ast}^{\rm eff,circ}$ as a function of the symmetric, dimensionless frequency parameter $x$, Eq. (\ref{6.24})
\begin{eqnarray}
 g_{S^\ast}^{\rm eff,circ}(x)&=&\frac32+\left(-\frac{9}{8}-\frac{3}{4}\nu\right) x+\left(-\frac{27}{16}-\frac{75}{8}\nu  \right) x^2\nonumber\\
&&+ \left[-\frac{405}{128}+\left(-\frac{7681}{192}+\frac{41}{32}\pi^2\right)\nu\right] x^3 \nonumber\\
&& + (q_4+q_4^{\ln{}}\ln x) x^4+  (q_5+q_5^{\ln{}}\ln x) x^5\nonumber\\
&&-\frac{93304}{1575}\pi\nu x^{11/2}\nonumber\\
&& + (q_6+q_6^{\ln{}}\ln x) x^6\nonumber\\
&& +\frac{4195411}{12600}\pi\nu x^{13/2}\nonumber\\
&& + (q_7+q_7^{\ln{}}\ln x+q_7^{\ln^2}\ln^2 x) x^7\nonumber\\
&&+\frac{5410915721}{34927200}\pi\nu x^{15/2}\,,
\end{eqnarray}
where
\begin{eqnarray}
q_4&=& -\frac{1456}{15}\nu\ln 2-\frac{1701}{256}\nonumber\\
&& +\left(-\frac{49069}{640}+\frac{25631}{2048}\pi^2-48\gamma \right)\nu\nonumber\\
q_4^{\ln{}} &=&-24 \nu \,,
\end{eqnarray}
\begin{eqnarray}
q_5&=& -\frac{15309}{1024}+\left(\frac{65656}{105}\ln(2)-\frac{729}{7}\ln(3)\right.\nonumber\\
&& \left.-\frac{162724753}{89600}+\frac{1415301}{8192}\pi^2+\frac{8992}{35}\gamma  \right)\nu \nonumber\\
q_5^{\ln{}} &=& \frac{4496}{35}\nu\,,
\end{eqnarray}
\begin{eqnarray}
q_6&=& -\frac{72171}{2048}+\left(-\frac{479758}{567}\ln 2 +\frac{152361}{140}\ln 3\right.\nonumber\\
&&  -\frac{1344552995}{7077888}\pi^2+\frac{702202}{2835}\gamma+\frac{16790137}{1048576}\pi^4\nonumber\\
&& \left. -\frac{42486972973176751}{23465490048000}\right)\nu\nonumber\\
q_6^{\ln{}} &=&\frac{351101}{2835}\nu \,,
\end{eqnarray}
\begin{eqnarray}
q_7&=& -\frac{2814669}{32768}+\left(-\frac{10458915881}{1455300} \gamma-\frac{21925633271}{3118500}\ln 2\right.\nonumber\\
&& -1088\zeta(3)-\frac{45020853}{12320}\ln 3-\frac{1953125}{3564}\ln 5\nonumber\\
&&  +\frac{58208}{105}\gamma^2+\frac{1167584}{525}\ln^2 2\nonumber\\
&& -\frac{200197499477}{335544320}\pi^4-\frac{13873084533949}{8918138880}\pi^2\nonumber\\
&& \left. +\frac{499904}{225}\gamma\ln 2+\frac{4364959541591800745880367}{53858368206010368000}\right)\nu\nonumber\\
q_7^{\ln{}} &=& \left(-\frac{10458915881}{2910600}+\frac{249952}{225}\ln 2+\frac{58208}{105}\gamma  \right)\nu \nonumber\\
q_7^{\ln^2{}} &=& \frac{14552}{105}\nu \,.
\end{eqnarray}
Let us finally note that Eq. (\ref{new_6.27}) above yields, when re-expanded in powers of $x$, the PN expansion of the function $m_2\Omega_1(x)$.

\section{Summary and concluding remarks}

We have indicated how Detweiler's redshift function $|k|(\Omega)$ could be extended into an infinite hierarchy of gauge-invariant functions associated with circular orbits.
Here, we focussed on the unique, one-derivative generalization of $|k|(\Omega)$, namely the function $|\nabla k|(\Omega)\equiv \sqrt{\frac12 \nabla_\mu k_\nu \nabla^\mu k^\nu}$. After discussing (in agreement with Ref. \cite{Dolan:2013roa}) both the kinematical (spin precession) and dynamical (spin-orbit coupling) significance of the function $|\nabla k|(\Omega)$, we obtained a simple explicit expression for $|\nabla k|$ in terms of the covariant components $g_{\mu\nu}^{(2+1)}$ of the equatorial reduction of the (regularized) metric, namely Eq. (\ref{2.11}). By expanding the latter expression to first order in the mass ratio $q=m_1/m_2\ll 1$, we derived an expression for the (gauge-invariant) $O(q)$ piece, $q\delta(y)$, Eq. (\ref{3.9}), in $|\nabla k|/\Omega$ (where $y=(m_2\Omega)^{2/3}$), in terms of the $O(q)$ piece, $q h_{\mu\nu}$, of the two-body metric, see Eq. (\ref{3.10}).
Using Regge-Wheeler-Zerilli-Mano-Susuki-Takasugi black hole perturbation technology, together with Barack-Ori-Hikida-Nakano-Sasaki mode-sum regularization, we succeded in analytically computing the post-Newtonian expansion of $\delta^{\rm R}(y)$, and $\delta \psi(y)=-\sqrt{1-3y}\delta^{\rm R} (y)$ up to the 8.5PN order, i.e., up to $O(y^{8.5})$ included, Eq. (\ref{4.33new}).
We then compared our analytic result to a recent numerical computation of $\delta\psi(y)$ by Dolan et al. \cite{Dolan:2013roa}.
We found that the successive PN approximants to $\delta \psi(y)$ exhibit a rather satisfactory \lq\lq convergence" towards the numerical data, and allow one to capture its most apparent strong-field features (a change of sign around $y\approx 0.17$, followed by a fast decrease towards negative values; see Fig. 1). By using results on the light-ring singular behavior of $h_{\mu\nu}^{\rm R}$ (in the Lorenz gauge) \cite{Akcay:2012ea}, we argued that $\delta \psi (y)$ will diverge as $\delta \psi (y)\approx -0.14 /(1-3y)$, Eqs. (\ref{5.9}) and (\ref{5.10}), as particle 1 approaches the light-ring ($r_0\to 3m_2$, $y\to 1/3$). It would be interesting to check this prediction numerically. We provided a simple, accurate global fit of the 1SF spin precession $\delta \psi(y)$ incorporating this pole-like behavior, together with the 3PN-level knowledge of $\delta \psi (y)$, see Eq. (\ref{5.11}).

We transcribed our kinematical spin-precession results (i.e., the higher order PN expansions of $\delta (y)$ and $\delta\psi (y)$), into a corresponding, high-order PN expansion of the second effective gyro-gravitomagnetic ratio $g_{S_\ast}^{\rm eff}(u)$ entering the spin-orbit part of the effective EOB Hamiltonian, see Eqs. (\ref{6.12}),  (\ref{6.14}), (\ref{6.16}).
Here, ${\mathbf S}_\ast$ refers to the second, basic symmetric combination, Eq. (\ref{6.6}), of the two spins ${\mathbf S}_1$ and ${\mathbf S}_2$, and $u$ denotes the gauge-invariant EOB gravitational potential $u=GM/(c^2R_{\rm EOB})$ (with $M=m_1+m_2$). [The function $g_{S_\ast}^{\rm eff}(u)$ is gauge-invariant and refers to a sequence of circular motions.]
We showed (see Eq. (\ref{our_res})) that the $O(\nu)$ piece (where $\nu=m_1 m_2/(m_1+m_2)^2$), $\nu g_{S_\ast 1\rm SF}^{\rm eff}(u)$, of $g_{S_\ast}^{\rm eff}(u)$ can be expressed as a linear combination of $\delta\psi(u)/u$, $a_{1\rm SF}(u)$ and $a'_{1\rm SF}(u)$, where 
$\nu a_{1\rm SF}(u)$ denotes the $O(\nu)$ piece in the basic, symmetric EOB radial potential $A(u; \nu)=1-2u+\nu a_{1\rm SF}(u)+O(\nu^2)$.
By combining our global fit for $\delta\psi(u)$ with a previously obtained global fit for $a_{1\rm SF}(u)$ \cite{Akcay:2012ea}, we obtained a global representation of the strong-field behavior of $g_{S_\ast 1\rm SF}^{\rm eff}(u)$, See Fig. 2. We also provided a simple, analytic fit for $g_{S_\ast 1\rm SF}^{\rm eff}(u)$, Eq. (\ref{gsstar_fit}).
A remarkable prediction of the global, strong-field knowledge of $g_{S_\ast 1\rm SF}^{\rm eff}(u)$ brought by our results, is that, while the $\delta\psi(u)/u$ contribution to $g_{S_\ast 1\rm SF}^{\rm eff}(u)$ would suggest, when considered by itself, a simple-pole plunge of $g_{S_\ast 1\rm SF}^{\rm eff}(u)$ towards $-\infty$ ($\propto - (1-3u)^{-1}$) as one approaches the light-ring ($u\to 1/3$), we found that the contributions depending on $a_{1\rm SF}(u)$, and especially $a_{1\rm SF}'(u)$, counterbalance this downwards plunge, and turn it into a stronger {\it upward} singular behavior near the light-ring of the form $g_{S_\ast 1\rm SF}^{\rm eff}(u)\approx + 0.094/(1-3u)^2$. As a consequence, we predict that $g_{S_\ast 1\rm SF}^{\rm eff}(u)$ (which is negative in the weak-field domain $u\ll 1$), will change its sign near $u\approx 0.28$, to become positive as $u\to 1/3$; see Fig. 2.
As the negative sign of the currently known PN expansion of  $g_{S_\ast 1\rm SF}^{\rm eff}(u)$ has played an important role in the various studies of the binding energy of spinning binaries \cite{Damour:2008qf,Barausse:2009xi}, our finding of such a strong-field sign change of 
$g_{S_\ast 1\rm SF}^{\rm eff}(u)$  might have important consequences for improving the current EOB-based modeling of the coalescence of spinning binaries \cite{Pan:2013rra,Taracchini:2013rva}.

Let us finally note that our work opens new research avenues, both for numerical GSF studies and for analytical ones. We already mentioned the importance of checking numerically the singular behavior of $\delta \psi(y)$ near the light-ring.
It would be quite useful (in view of the EOB based modeling of coalescing binary neutron stars \cite{Damour:2009wj,Baiotti:2010xh,Bini:2012gu,Bernuzzi:2012ci})
to numerically compute the two-derivative gauge-invariant functions mentioned in the Introduction, and notably the 1SF piece of the electric-quadrupole tidal invariant ${\mathcal E}^2(\Omega)$. We intend to apply our analytic approach to a computation of the PN expansion of the latter function. GSF computations of other gauge-invariant functions (involving higher derivatives of $k$, and/or of the curvature) might also provide important information, especially if performed in presence of a nonzero spin ${\mathbf S}_2$ of the large mass $m_2$. 
[Note that, from the point of view of PN regularization theory (using dimensional continuation) \cite{Damour:1982wm,Damour:2001bu,Blanchet:2003gy} the gauge-invariant functions $|k|(\Omega)$, $|\nabla k|(\Omega)$, ${\mathcal E}^2(\Omega)$, $\ldots$ are defined to all orders in the mass ratio: see Refs. \cite{Blanchet:2009sd,Bini:2012gu,Bohe:2012mr}.]
Clearly, there are here many possibilities for fruitful synergies between PN theory, GSF theory and EOB theory.

\noindent {\bf Acknowledgments.}  
D.B. thanks the Italian INFN (Naples) for partial support and IHES for hospitality during the development of this project.
Both  authors are grateful to ICRANet for partial support.

\end{document}